\documentclass{aa}  

\usepackage{graphicx}
\usepackage{txfonts}
\usepackage{float}
\usepackage{enumitem}
\usepackage{ctable}
\begin{document}

\title{Constraining the turbulence and the dust disk in IM~Lup: onset of planetesimal formation}

\author{Riccardo Franceschi \inst{1}, Tilman Birnstiel\inst{2,3}, Thomas Henning\inst{1}, Anirudh Sharma\inst{2}
}

\institute{Max-Planck-Institut f\"ur Astronomie (MPIA),
    K\"onigstuhl 17,  69117 Heidelberg, Germany
    \and University Observatory, Faculty of Physics, Ludwig Maximilians University, Scheinerstr. 1, 81679 Munich, Germany
    \and Exzellenzcluster ORIGINS, Boltzmannstr. 2, D-85748 Garching, Germany
    \email{franceschi@mpia.de}
}


\abstract
{Observations of protoplanetary disks provide information on planet formation and the reasons for the diversity of planetary systems. The key to understanding planet formation is the study of dust evolution from small grains to pebbles. Smaller grains ($\sim 1 \;\mu m$) are well-coupled to the gas dynamics, and their distribution is significantly extended above the disk midplane. Larger grains settle much faster and are efficiently formed only in the midplane. By combining near-infrared polarized light and millimeter observations, it is possible to constrain the spatial distribution of both the small and large grains.}
{We aim to construct detailed models of the size distribution and vertical/radial structure of the dust particles in protoplanetary disks based on observational data. In particular, we are interested in recovering the dust distribution in the IM~Lup protoplanetary disk.}
{We create a physical model for the dust distribution of protoplanetary disks. We then simulate the radiative transfer of the millimeter continuum and the near-infrared polarized radiation. Using a Markov chain Monte Carlo method, we compare the derived images to the observations available for the IM~Lup disk to constrain the best physical model for IM~Lup and to recover the vertical grain size distribution.}
{The millimeter and near-infrared emission tightly constrain the dust mass and grain size distribution of our model. We find size segregation in the dust distribution, with millimeter-sized grains in the disk midplane. These grains are efficiently formed in the disk, possibly by sedimentation-driven coagulation, in accord with the short settling timescales predicted by our model. This also suggests a high dust-to-gas ratio at smaller radii in the midplane, possibly triggering streaming instabilities and planetesimal formation in the inner disk. We obtain a turbulent $\alpha$ parameter of $3 \times 10^{-3}.$}
{}


\authorrunning{R. Franceschi \inst{1}, T. Birnstiel \inst{2,3}, Th. Henning \inst{1}
}

\institute{Max-Planck-Institut f\"ur Astronomie (MPIA),
    K\"onigstuhl 17,  69117 Heidelberg, Germany
    \and University Observatory, Faculty of Physics, Ludwig Maximilians University, Scheinerstr. 1, 81679 Munich, Germany
    \and Exzellenzcluster ORIGINS, Boltzmannstr. 2, D-85748 Garching, Germany
    \email{franceschi@mpia.de}
}
\titlerunning{PPDs mass determination from grain evolution}
\authorrunning{Franceschi et al.}
\maketitle
%

\section{Introduction}
Protoplanetary disks around young stars are the nursery of planetary systems. These disks are almost entirely composed of gas and a small mass fraction of dust. Small dust grains ($\sim 1 \; \mu m$) are well-coupled to the gas structure, and therefore extend vertically over several scale heights, following the gas distribution. Dust settling in turbulent disks was extensively studied (e.g. \citealt{Dubrulle95, Schrapler04, Fromang09, Zsom11, Woitke16}). On the contrary, large millimeter grains settle much faster towards the disk midplane, where the growth process is particularly efficient because of the high densities. By constraining the distribution of both large and small dust grains, we can extrapolate the grain's vertical size distribution, probing disk fundamental properties such as the turbulent structure and the dust-to-gas ratio.

Disks tend to have a flared structure \citep{Kenyon87, Bell97, Chiang97}. Flaring occurs when the midplane temperature falls off more slowly than $r^{-1}$. The disk surface then flares outward with increasing radius as a consequence of vertical hydrostatic equilibrium. With the development of our observational techniques, we characterized several examples of disks with a flared structure (e.g. \citealt{Pinte16, Avenhaus18,  Villenave20, Law22}). For instance, \cite{Law22} recently observed a correlation between the CO emitting height and the disk size in a sample of disks observed with the Atacama Large Millimeter/submillimeter Array (ALMA). Using the dust emission, \cite{Avenhaus18} measured the flaring of V4046~Sgr, RXJ~1615, and IM~Lup with the Very Large Telescope (VLT) Spectro-Polarimetric High-contrast Exoplanet REsearch (SPHERE) instrument. These near-infrared polarized light images show unprecedented details of the disk's surface layer by tracing the scattering of small, micron-sized dust grains. At these wavelengths, the disk is expected to be optically thick and the observed light is dominated by light scattered by the dust grains on the disk surface. Among these disks, IM~Lup observations are arguably among the most impressive images of a protoplanetary disk. IM~Lup is part of the Disks ARound TTauri Stars with SPHERE (DARTTS-S) survey by \cite{Avenhaus18}, targeting eight TTauri stars, at $1.25 \; \mu m$ and $1.65 \; \mu m$ wavelengths. With its intermediate inclination ($\sim 56^\circ$), we have a three-dimensional perspective of the disk around IM~Lup, which can be used to constrain the vertical disk structure and the small grain distribution in a model-independent way.

The large dust grain properties can be similarly constrained through observations. At (sub-)millimeter wavelengths, ALMA data provided a new understanding of the properties and distribution of large, millimeter-sized dust grains, found in the disk midplane (e.g. \citealt{Andrews16, Huang17, Pinte18}). IM~Lup was also part of the Disk Substructures at High Angular Resolution Project (DSHARP) survey (\citealt{Andrews18} and following papers), conducted with ALMA. The survey led to the characterization of substructures for 20 nearby protoplanetary disks using observations at 1.25~mm (240~GHz).

By combining ALMA and SPHERE images, it is possible to further characterize the structure of disks, and how they affect the planet formation processes. The SPHERE and ALMA data show two very different geometries of the disk. In the near-infrared, the disk shows strong flaring and a prominent vertical structure. This implies that small grains are well coupled to the gas structure, are located in the upper layers of the disk, and are prevented from settling \citep{Avenhaus18}. The millimeter image, on the other hand, shows a flat disk which suggests that the larger dust grains are well settled to the midplane. The disk emission differs also in its radial extent: the ALMA data extend up to 290~au, while the SPHERE data go as far as 400~au. These two images can be used to model the distribution of the largest and smallest dust grains, which we can then use to constrain the full grain size distribution.

\cite{Pinte08} took a similar approach, and combined infrared spectra and scattered light emission with millimeter thermal emission to constrain the parameter range of their IM~Lup model. The data consists of scattered light detected with the WFPC2 instrument on board of the Hubble Space Telescope (HST), Spitzer near and mid-infrared spectroscopy and SubMillimeter Array (SMA) millimeter emission. They combined a manual exploration of the model parameter space with a more rigorous fitting of the parameters that cannot be directly inferred from the observations. They found IM~Lup to be a massive disk, with a mass of about $\sim 0.1$ M$_\odot$  assuming  a $1\%$ dust-to-gas ratio, extending up to $\sim 400$~au, but with a low H$\alpha$ emission, hinting at a low accretion rate. However, a low accretion rate is surprising for a massive disk, and they suggest that may be due to a period of low or moderate accretion. More recently, the accretion rate was measured by \cite{Alcala17, Alcala19} using the spectrum observed with the VLT/X-Shooter spectrograph and the far-utraviolet continuum excess emission observed with the HST-Cosmic Origins Spectrograph (HST-COS) and -Space Telescope Imaging Spectrograph (HST-STIS). These studies report a high mass-accretion rate of $10^{-8} \; M_\odot$. \cite{Pinte08} found that a dust population perfectly mixed with the gas cannot explain both the dust infrared spectral features and the millimeter continuum. Millimeter-sized  grains must be present in the disk, but their emission cannot come from the same region as the infrared emission. The lower modeling power and quality of the data available at the time of this study, however, did not allow to uniquely constrain the stratified dust disk structure, and other disk models than their best fit model can match the used data.

In this paper, we introduce a comprehensive modeling of protoplanetary disk dust distribution, focusing in particular on the vertical distribution of small grains. We then test this model by fitting it to the high-resolution millimeter ALMA data of IM~Lup and near-infrared polarized radiation SPHERE data using a Markov chain Monte Carlo (MCMC) algorithm. This allows us to constrain the complete dust grain size distribution of the disk. Finally, we discuss our results and how they affect our understanding of the dust structure of protoplanetary disks.

\section{Disk model}
\label{sec:model}
The gas distribution for our disk physical model is the axisymmmetric, self-similar Lynden-Bell \& Pringle profile \citep{LyndenBell74}:
\begin{equation}
    \label{lynden bell pringle}
    \Sigma(r) = \Sigma_c \left( \frac{r}{r_c} \right)^{-\gamma} \exp \left[ \left( -\frac{r}{r_c} \right)^{2 - \gamma} \right],
\end{equation}

where $r_c$ is the characteristic radius and $\gamma$ is the gas surface density exponent. To recover the vertical gas distribution, we need to compute the vertical scale height of the disk. This scale height is expressed by the ratio of the sound speed $c_s$ to the Keplerian angular velocity $\Omega$:

\begin{equation}
    \label{eq:pressure scale height}
    H_p(r) = \frac{c_s}{\Omega} = \left( \frac{k \; T_{mid}(r) \; r^3}{\mu \; m_H \; G \; M_\star} \right)^{1/2},
\end{equation}

where $k$ is the Boltzmann constant, $T_{mid}(r)$ the midplane gas temperature distribution, $\mu = 2.3$ the reduced mass of the $H_2$ molecule, $m_H$ the hydrogen atomic mass, G the gravitational constant, and $M_\star$ the stellar mass. This equation comes from balancing the vertical pressure force of the gas with the gravitational force towards the midplane. Using this scale height, we obtain the gas density at a radius $r$ and height $z$:

\begin{equation}
    \label{eq:hydrostatic_equilibrium}
    \rho (r, z) = \frac{\Sigma(r)}{\sqrt{2\pi} H_p(r)} \exp \left( -\frac{1}{2} \left( \frac{z}{H_p(r)} \right)^2 \right).
\end{equation}

The two terms $H_p(r)$ and $T_{mid}(r)$ are coupled with each other: the amount of starlight penetrating to the disk midplane is determined by $H_p$. This radiation increases the midplane temperature $T_{mid}$, which results in a higher pressure scale height. Therefore, $H_p$ and $T_{mid}$ need to be consistent with each other, and the disk structure must be computed iteratively. To estimate the amount of heating from stellar irradiation, we need to compute the optical depth as a function of the location in the disk and the flaring angle $\varphi(r)$, the angle at which the stellar radiation hits the disk surface. The flaring angle is, according to \cite{Chiang97}:

\begin{equation}
    \label{eq:flaring index}
    \varphi(r) = \frac{0.4 R_\star}{r} + r \frac{d}{dr} \left( \frac{H_s}{r} \right),
\end{equation}

where $H_s$ is the surface height of the disk. Note that this is a different quantity than the pressure scale height $H_p$ (which is usually lower), and is defined by the height above the midplane where the optical depth to the stellar radiation is unity.

Given the irradiation angle $\varphi(r)$, which is assumed to not depend on $z$, the optical depth to the stellar radiation is:
\begin{equation}
    \tau_\star(r, z) = \frac{\kappa_{P, *}}{\varphi(r)} \int_z^{\infty} \varepsilon \; \rho_g(r, z')  dz',
\end{equation}
where $\kappa_{P, \star}$ is the Planck mean opacity at stellar wavelengths, $\varepsilon$ is the assumed dust-to-gas ratio, and $\rho_g$ is the gas density.

We now wish to find the value of $z$, at a given $r$, at which $\tau_\star = 1$. \cite{Dullemond01} found that this is equivalent to solving:

\begin{equation}
    1 - \mathrm{erf} \left( \frac{H_s}{\sqrt{2} \; H_p} \right) = \frac{2 \; \varphi}{\tau_\star},
\end{equation}

where erf is the error function:

\begin{equation}
    \mathrm{erf} \; z = \frac{2}{\sqrt{\pi}} \int_0^z e^{-t^2} dt.
\end{equation}

This estimate for $H_s$ can now be used to derive the flaring index from Eq.\ref{eq:flaring index}. Since the flaring angle determines $T_{mid}$, which determines $H_p$, these equations can be solved iteratively to derive $H_p$, $H_s$, $\varphi$, and $T_{mid}$ consistently.

\subsection{Vertical structure}
To calculate the irradiation of the disk by the star in a 2D model, we use a ray-tracing technique, a 1+1D approach \citep{Dullemond02}. This allows us to approximate a full 2D solution by calculating the transport of the stellar radiation through the disk, even though there is no radial transfer of thermal energy in the gas. This is a good approximation of the disk structure  achievable within a reasonable computation time. Computational time is a strong constraint to our model: each new set of parameters we test against the observational data requires the calculation of a new disk structure, and new radiative transfer calculations to produce the images to compare to the observations. A faster disk model allows us to test more parameters with the same computational resources, and a better sampling of our posterior parameters distribution.

Ray-tracing consists of a simple integration on the lines $\zeta = z/r = const$. The stellar flux can then rewritten as:

\begin{equation}
    \label{eq:stellar_flux}
    F_\star (r, \zeta) = \frac{L_\star}{4\pi \; (r^2 + z^2)} e^{-\tau_\star(r, \zeta)},
\end{equation}

and the optical depth as:

\begin{equation}
    \tau_\star = \sqrt{1 + \zeta^2} \int_{R_\star}^r \rho_g (r',\zeta) \; \kappa_{P,\star} \; dr'
\end{equation}.

From this radiation field, we can compute the temperature as:

\begin{equation}
    \label{eq:temperature_profile}
    T(r, \zeta) = \left[
        \frac{\pi}{\sigma_{SB}} \left(
        J_{diff}(r, \zeta)
        + \frac{\kappa_{P \star}}{\kappa_{P}(T)} J_\star(r, \zeta)
        \right)
        \right]^{1/4},
\end{equation}

where $\sigma_{SB}$ is the Stephan-Boltzmann constant, $\kappa_{P \star}$ the mean Planck opacity at stellar wavelength, and $\kappa_P(T)$ is the mean Planck opacity of the dust at temperature $T$. $J_{diff}$ and $J_\star$ are the mean intensity from dust re-emission and the intensity from stellar radiation, which can be derived from Eq.\ref{eq:stellar_flux} and the dust opacity, which is discussed in Sec.\ref{sec:opacity}. Once we have the temperature profile, we can solve for the vertical density profile from Eq.\ref{eq:hydrostatic_equilibrium}. This will be a different profile from the one used to solve Eq.\ref{eq:temperature_profile}, so we iterate these solutions until convergence.


\subsection{Dust}
\label{sec:dust}
We already discussed that ALMA and SPHERE observations indicate that the disk around IM~Lup is composed of at least two dust populations: one of small grains, coupled to the gas structure and vertically extended, and one of large grains settled to the midplane. These are just two parts of a continuous size distribution of grains, as predicted by dust coagulation models (e.g. \citealt{Dullemond05, Brauer08, Birnstiel10, Birnstiel12}). Since a detailed dust coagulation/fragmentation model is computationally expensive, parametric functions to model the dust grain distribution are common in the literature.

We assume that the largest grain size at each radial location is given by a power-law distribution:

\begin{equation}
    \label{eq:a_max}
    a_{max}(r) = a_0 \; \left( \frac{r}{r_c} \right)^{-\gamma},
\end{equation}

where $a_0$ is the maximum grain size at radius $r_c$. From this maximum grain size at any radial location we can reconstruct the grain size distribution using:

\begin{equation}
    \label{eq:size_distribution}
    n(a) \propto \left\{
    \begin{array}{ll}
        a^{-q} & \textrm{for $a_{min} \leq a \leq a_{max}$}, \\
        0      & \textrm{else},
    \end{array}
    \right.
\end{equation}

where $a_{min}$ is $10^{-5}$ cm, $a_{max}$ is given by Eq.\ref{eq:a_max}, and $n(a)$ is normalized to the total dust volume density:

\begin{equation}
    \label{eq:total_dust_density}
    \rho_d = \int_{0}^{\infty }n(a) \; m(a) \; da,
\end{equation}

with $m(a)$ the mass of a grain of radius $a$. The value of the index q in Eq.\ref{eq:size_distribution}, one of our model parameters, has already been investigated in previous studies. Measurements of the interstellar extinction (e.g. \citealt{Mathis77}) found $q \approx 3.5$, which is also consistent with more recent submillimeter observations of debris disks \citep{Ricci15}. However, the physics of gas-rich young protoplanetary disk is very different from the one of gas-depleted debris disks. \cite{Birnstiel11} used dust coagulation and fragmentation models to fit an analytic multi-power law to dust evolution models.

The total dust surface density is derived from the gas density using a dust-to-gas ratio:

\begin{equation}
    \label{eq:d2g}
    \varepsilon(r) = \varepsilon_0 \left( \frac{r}{r_0} \right)^{-p},
\end{equation}

where $\varepsilon_0$ is the dust-to-gas ratio at the normalization radius $r_0$.

From this estimate of the dust density, we compute the vertical dust distribution by balancing turbulent diffusion and vertical settling. Assuming steady state, this is equivalent to solving \citep{Dubrulle95, Schrapler04, Fromang09}:

\begin{equation}
    \label{eq:dust_vertical}
    \frac{\partial}{\partial z} \left( \log \frac{\rho_d}{\rho} \right) = - \frac{\Omega^2 \; \tau_s}{D} \; z,
\end{equation}

where $\tau_s$ is the grains stopping time, $D$ is the turbulent diffusivity, and $\Omega$ the Keplerian angular velocity. A grain stopping time is the typical time it takes for a grain initially at rest to reach the local gas velocity. This depends on the grain bulk density $\rho_s$ and its size $a$:

\begin{equation}
    \label{eq:stopping_time}
    \tau_s = \frac{\rho_s \; a}{\rho \; c_s},
\end{equation}

where $c_s$ is the local sound speed. Different grain populations will have different vertical distributions, as larger grains quickly settle toward the midplane, while small grains remain coupled to the vertical structure of the gas. In our models, we divide the grain distribution from Eq.\ref{eq:size_distribution} in 30 populations of different grain sizes, logarithmically spaced between $a_{min}$ and $a_{max}$, and solve their vertical distribution, Eq.\ref{eq:dust_vertical}.

The turbulent diffusivity $D$ is a parametrization of how the grains are removed from the midplane by turbulent velocity fluctuations of the gas. The simplest case is to assume that is constant:

\begin{equation}
    \label{eq:diffusivity}
    D = \frac{\alpha \; c_s \; H_p}{\mathit{Sc}},
\end{equation}

where $\mathit{Sc}$ is the Schmidt number, which has been measured to be of order one in zero net MHD turbulence \citep{Johansen05, Johansen06}. The diffusivity is intimately linked to the turbulence properties of the gas flows, therefore assuming it to be constant in space is a reasonable assumption when the turbulence is homogeneous. The most likely deviation from homogeneous turbulence is MHD effects, which are vertically and radially stratified \citep{Dzyurkevich13}. As a consequence, the diffusivity coefficient $D$ will change with the disk height $z$ at any given radius. In our models, we do not explore MHD effects on the disk structure, and we use a constant diffusivity as in Eq.\ref{eq:diffusivity}

\subsection{Dust opacity}
\label{sec:opacity}
To compare our models to the observations we produce synthetic emission maps of the dust structure. It requires us to calculate the opacity of the dust distribution derived in Sec.\ref{sec:dust}. We follow the approach of \cite{Birnstiel18}. This work analyzed the millimeter emission of DSHARP disks (including IM~Lup) to study the dust properties of these sources.

The grain opacity is strongly affected by the grain composition. Here we adopt the same composition used in the DSHARP project \citep{Birnstiel18}, a mixture of water ice, astronomical silicates, troilite, and refractory organic material (see Tab.\ref{table:dust_composition}). \cite{Birnstiel18} assumed particles without porosity, however grains are expected to have larger porosities, at least in the initial stage of collisional growth (e.g. \citealt{Krijt15}), therefore we assume grains of 30\% porosity.

\begin{figure}
    \centering
    \includegraphics[width=8cm]{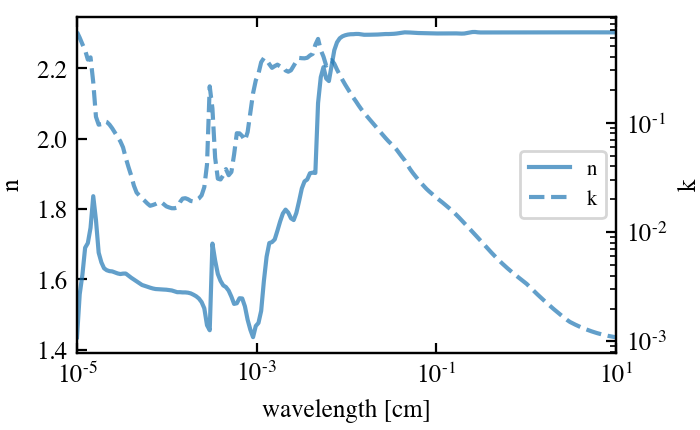}
    \caption{Medium optical constant for the dust composition in     Tab.\ref{table:dust_composition} \citep{Birnstiel18}}
    \label{fig:optical_constants}
\end{figure}

To calculate the dust mass absorption and scattering coefficients $\kappa^{abs}_\nu$ and $\kappa^{sca}_\nu$ as a function of frequency $\nu$, we need the optical constants of the composing materials $n(\lambda)$ and $k(\lambda)$, which depend on the observed wavelength $\lambda$. In Tab.\ref{table:dust_composition} we report the dust composition assumed and the laboratory experiments providing the optical constants of the composing materials. Deriving the optical constants of a medium composed of multiple materials is a challenging task, as it depends on how we assume these materials are distributed in the dust grains. For general cases some computationally expensive numerical models need to be used, but in some limit cases analytic solutions are possible. For a homogeneous mix of the  grain components, the solution is given by the Bruggemann rule (see \citealt{Bohren98} for more details). The resulting optical constants are shown in Fig.\ref{fig:optical_constants}.

\begin{table*}
    \caption{DSHARP dust composition, from \cite{Birnstiel18}}
    \label{table:dust_composition}
    \centering
    \begin{tabular}{c c c c c}
        \hline\hline

        material               & reference        & bulk density & mass fraction & volume fraction \\
                               &                  & [g/cm$^3$]   &               &                 \\
        \hline
        water ice              & \cite{Warren08}  & 0.92         & 0.200         & 0.3642          \\
        astronomical silicates & \cite{Draine03}  & 3.30         & 0.3291        & 0.1670          \\
        troilite               & \cite{Henning96} & 4.83         & 0.0743        & 0.0258          \\
        refractory organics    & \cite{Henning96} & 1.50         & 0.3966        & 0.4430          \\
        \hline
    \end{tabular}
    \tablefoot{The bulk density of the mix is $\rho_2 = 1.675$~g/cm$^3$}
\end{table*}

To calculate the size dependent opacities $\kappa_\nu^{abs}$ and $\kappa_\nu^{sca}$ from the optical constants in Fig.\ref{fig:optical_constants} we use the \texttt{dsharp\_opac}\footnote{https://github.com/birnstiel/dsharp\_opac} code, introduced by \cite{Birnstiel18}, based on the original Mie calculation code by \cite{Bohren98}.

To derive the total absorption opacity we create a grid of 30 grain sizes, logarithmically spaced between $10^{-5}$ and $10$~cm. The opacity $\kappa_\nu^{abs,; tot}$ (Eq.\ref{eq:size_distribution}) at frequency $\nu$ is then averaged over the grain distribution $n(a)$ (Eq.\ref{eq:size_distribution}):

\begin{equation}
    \label{eq:total_absorption}
    \kappa_{\nu}^{abs, \;  tot} =  \frac{\int_{a_{min}}^{a_{max}} n(a) \; m(a) \; \kappa_{abs}^{sca}(a) \; da}{\int_{a_{min}}^{a_{max}} n(a) \; m(a) \; da},
\end{equation}

where $a_{min} = 10^{-5}$~cm and $a_{max}$ is taken from Eq.\ref{eq:a_max}. Similarly, we calculate the total scattering opacity $\kappa_{\nu}^{sca, \; tot}$. Finally, to predict the map of the dust continuum and scattered light emission from the model described above using these opacities, we simulate the three-dimensional radiative transfer with the Monte Carlo radiative transfer code \texttt{RADMC-3D}\footnote{https://www.ita.uni-heidelberg.de/\textasciitilde dullemond/software/radmc-3d/} \citep{Dullemond12}

\section{Observations}
\label{sec:observations}
To constrain the vertical dust grain height in a disk we need a disk meeting three requirements. First, the disk must be moderately inclined ($\sim 30-70^\circ$), so that its three-dimensional structure can be reconstructed from its two-dimensional projection on the sky. Secondly, it must show some ring structures in the scattered light emission to estimate the vertical extension of small grains as a function of radius. Finally, we require millimeter continuum observations to constrain the large grain distribution in the disk midplane. The faint, smooth, and extended emission observed in the scattered light in the disk around IM~Lup makes this disk the ideal candidate for a test study. The methods developed here can then be applied in a future work to other axisymmetric disks and extended to include the effects of radial structures on the radial grain distribution due to mixing and drift using methods similar to the ones presented in this work for the vertical grain distribution.

The protoplanetary disk around IM~Lupus has been extensively studied in the  infrared scattered light \citep{Pinte08, Stapelfeldt14, Avenhaus18}, millimeter continuum \citep{Pinte08, Cleeves16, Andrews18, Pinte18}, and gas line emission \citep{Cleeves16, Pinte16}. In particular, \cite{Zhang21} used high-resolution ALMA observations to study CO isotopologue emission, and retrieved the gas mass and distribution of several disks, including IM~Lup. In our models, we assume the gas distribution derived in this work. This disk is the best candidate to test the simultaneous modeling of large and small grains, with the bonus of having extensive models of its gas structure to check the consistency of our results. Indeed, the efficient coupling of small grains to the gas implies similarities in their spatial distribution. To constrain our models, we use two sets of observations. One is the SPHERE measurements of the $\mathrm{1.65 \; \mu m}$ near-infrared polarized emission, tracing the scattering of stellar radiation by small grains on the disk surface \citep{Avenhaus18}. The second is the 1.25~mm continuum emission from the large grains in the disk midplane, detected by ALMA as part of the DSHARP disk survey \citep{Andrews18}. A summary of the parameters of IM~Lup can be found in Tab.\ref{table:IMLup_param}

\begin{table*}[]
    \caption{IM Lupi stellar and disk properties}
    \label{table:IMLup_param}
    \centering
    \begin{tabular}{c c c c c c c c}
        \hline\hline
        inc                            & PA                               & d                                 & age                                & $\mathrm{M _{disk}}$    & $\mathrm{T_{eff}}$      & $\mathrm{L_\star}$      & $\mathrm{M_\star}$      \\
        $\mathrm{\left[deg\right]}$    & [deg]                            & [pc]                              & [Myr]                              & $\mathrm{[M_\odot]}$    & [K]                     & $\mathrm{[L_\odot]}$    & $\mathrm{M_\odot}$      \\
        \hline
        $56 \pm 2$ \tablefootmark{(1)} & $145 \pm 2$  \tablefootmark{(1)} & $158 \pm 0.5$ \tablefootmark{(2)} & $1.1 \pm 0.2 $ \tablefootmark{(1)} & 0.2 \tablefootmark{(3)} & 4350  \tablefootmark{(4)} & 2.6 \tablefootmark{(4)} & 1.1 \tablefootmark{(3)} \\
        \hline
    \end{tabular}
    \tablefoot{
        References are: 1. \cite{Avenhaus18}; 2. \cite{GAIA16}; 3. \cite{Zhang21}; 4. \cite{Alcala19}
    }
\end{table*}

There are two possible approaches to compare our model images to the observations: either we directly compare the model images to the observed images, or we compare their radial emission profiles. We choose the latter, to ensure we give the same weight to both the millimeter and near-infrared data when constraining the models. Indeed, a direct comparison of disk models to data from different instruments is a challenging task, especially in this case where the image resolution element is defined differently (beams for ALMA data and pixels for SPHERE data). On the other hand, using azimuthally averaged brightness profiles allows us to avoid the subtleties of working directly on the imaging data. Since either approach is valid and does not bring more benefit than the other, we choose to work with the azimuthally averaged profiles to simplify our analysis.

In addition, we constrain our models using only the data outside of 1~arcsec (158~au) radius. We are primarily interested in characterizing the difference between the small and large grain distribution, so we focus on the outer regions of the disk where this difference is maximized. Secondly, the SPHERE coronograph is obscuring the inner 0.1~arcsec ($\sim$ 16~au) of the disk , making the observed inner disk always dimmer than in the models. Finally, IM~Lup observations show evidences both in the large grain \citep{Huang18-DSHARPII} and small grain emission \citep{Avenhaus18} up to 1~arcsecond. Here we do not address the effect these structures have on the dust distribution, which will be addressed in a follow-up study, and we focus on the region outside the 1~arcsec radius. 

\subsection{Near-infrared scattered light: height of small grain distribution}
\label{sec:observations_sphere}
\begin{figure}
    \centering
    \includegraphics[width=9cm]{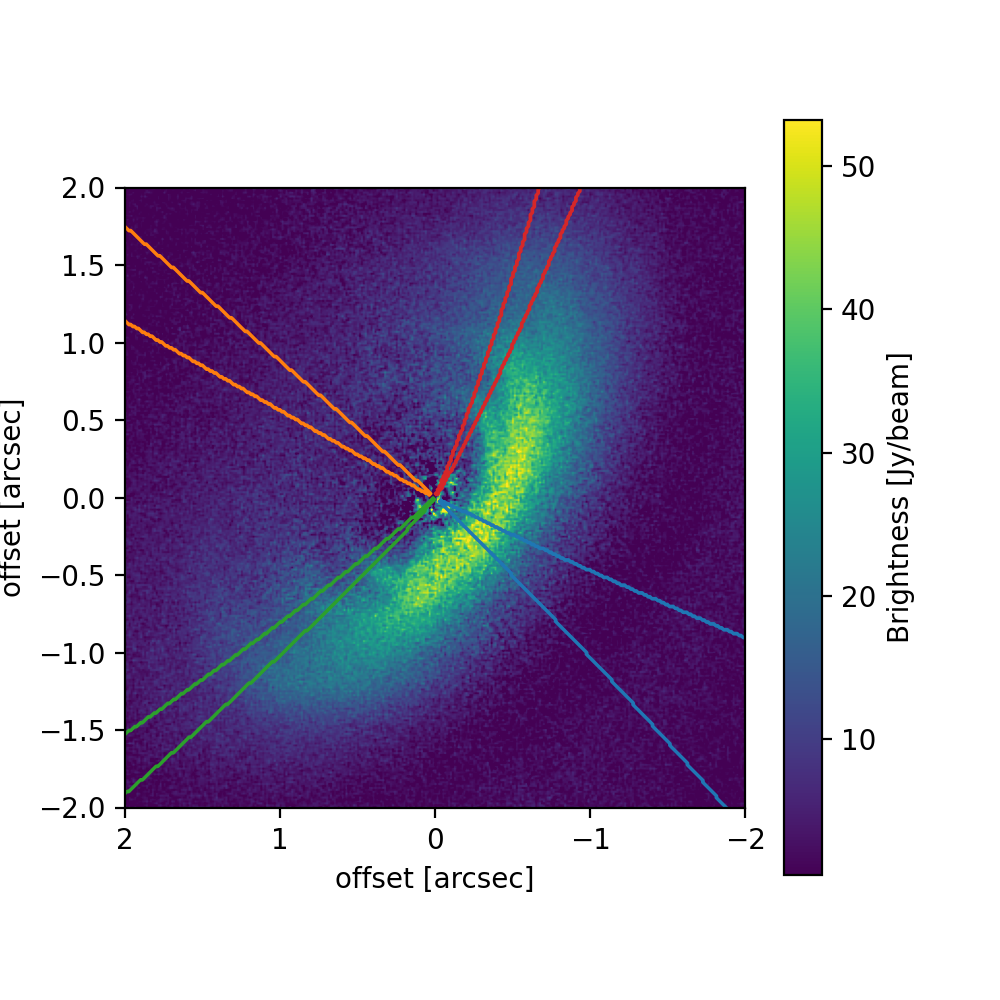}
    \caption{IM~Lup polarized emission at $1.65 \; \mu m$ wavelength observed by SPHERE \citep{Avenhaus18}. The cones show the regions over which we extract the radial profiles we use to compare our models.
    }
    \label{fig:obs_sca}
\end{figure}

The disk around IM~Lup has been observed in its scattered light emission by the Hubble Space Telescope (HST) \citep{Pinte08} and SPHERE \citep{Avenhaus18} (shown in. Fig.\ref{fig:obs_sca}. Scattered light images of an inclined disk such as IM~Lup are expected to show azimuthal asymmetries due to the anisotropic nature of scattering by dust grains. These anisotropies are caused by both grain properties and the flared disk geometry. Grain scattering favors scattering in the forward direction, and the disk region closer to the observer is the brightest (e.g. \citealt{Mulders13}). Moreover, the fraction of light that gets polarized by grain scattering depends on the grain structure, composition, and scattering angle.

The disk geometry also affects how much light is scattered by the disk surface. These images trace stellar radiation scattered by the small grains near the disk surface, because the disk is optically-thick at infrared wavelengths (e.g., \citealt{Grady00, Grady09, Hashimoto11, Monnier17}), and a flared disk geometry can cause an excess in its infrared emission.

IM~Lup scattered light emission has its outer edge at $\approx 340$~au, and a faint halo extending up to $\approx 700$~au \citep{Avenhaus18}. The SPHERE data reveal several ring structures. If we assume these rings to be circular and centered on the star, we can fit ellipses to the peak flux of the rings to estimate radius, inclination, position angle, and the vertical displacement from the midplane due to the height of the small grains. These measurements are carried out in \cite{Avenhaus18}, reported also in Tab.\ref{tab:IMLup_rings} for convenience.

\begin{table}[]
    \centering
    \begin{tabular}{c c c}
        \hline\hline
        radius           & flaring         & height          \\
        $\mathrm{[au]}$  &                 & [au]            \\
        \hline
        $91.9 \pm 3.2$   & $0.18 \pm 0.03$ & $16.5 \pm 3.3$  \\
        $152.1 \pm 4.8$  & $0.18 \pm 0.04$ & $27.4 \pm 6.9$  \\
        $240.8 \pm 4.8$  & $0.23 \pm 0.04$ & $55.4 \pm 10.7$ \\
        $332.8 \pm 12.7$ & $0.25 \pm 0.05$ & $83.2 \pm 19.8$ \\
        \hline
    \end{tabular}
    \caption{IM~Lup small grain ring parameters, as in \cite{Avenhaus18}}
    \label{tab:IMLup_rings}
\end{table}

They find that the dust scattering surface is described by a power law in radius:

\begin{equation}
    \label{eq:dust_scale_height}
    H_s(r) = H_0 \left( \frac{r}{r_0} \right)^\beta,
\end{equation}

where $H_0$ is the dust scale height at the fiducial radius $r_0$, and $\beta$ the flaring index. The height of the scattering surface is typically about 3-4 times the pressure scale height.

We use these results to extrapolate the radial scattered light emission profile of IM~Lup. The flaring index gives us the information necessary to reconstruct the disk 3D structure from its projected 2D image on the sky plane, the SPHERE data. This can be done using the GoFish package \citep{GoFish}. To preserve the information about the non-isotropic nature of the polarized scattered light emission, we extract the emission profiles in four different directions. As shown in Fig.\ref{fig:obs_sca}, we choose four cones $10^\circ$ wide in  PA in the forward and backward scattering direction, and the two side directions $90^\circ$ away. We apply this same procedure to the simulated disks, then we compare models to observations by fitting these four profiles.

\subsection{Millimeter continuum emission: large grains in the disk midplane}
\label{sec:millimeter observations}
The second set of observational constraints is the high resolution ($\sim 5$~au) millimeter continuum emission from large grains in the disk midplane, at $\lambda_{obs} = 1.25$~mm. These data are from the ALMA DSHARP program \citep{Andrews18}. Since both our model and the continuum emission are azimuthally axisymmetric, we compare our models to the azimuthal average of the observations. As with the scattered light emission, the millimeter emission shows radial structures in the inner disk, a spiral pattern extending from 25~au to 110~au \citep{Huang18}. Outside of this region, there is a gap in the emission profile. For this reason, we constrain our models with the observed emission profile outside the 1~arcsec radius, as with the scattered light data. The 1.25~mm emission has a sharp truncation at 295~au, also found in previous studies \citep{Pinte18}, and extends up to about 400~au.




\cite{Tazzari21} analyzed the 0.9~mm, 1.3~mm, and 3.1~mm emission for the 26 brightest disks in the Lupus region, including IM~Lup. One of their most interesting results is that they measure very little variation in disk size across different wavelengths. This behavior is common for highly structured disks, such as IM~Lup and the other DSHARP sources. For instance, it was also described by \cite{Pinilla15, Pinilla21}, which studies the effect of disk radial structures on the dust grain size distribution. The most obvious explanation would be a physical drop in the dust density, but an emission drop can also be explained by an opacity feature. For example, \cite{Rosotti19} and \cite{Powell17} argued that the outer edge of the emission profile at a given millimeter wavelength is also the location where the maximum grain size corresponds to the opacity resonance at that wavelength ($a_{max} \sim \lambda / 2 \pi$). When the maximum grain size, decreasing with radius, falls below this size, the grain opacity drops quickly, and we observe a sharp truncation in the emission profile.  This is in accord with the predictions of dust growth models and radial drift (e.g. \citealt{Birnstiel16, Powell19, Franceschi22}). However, if this truncation were caused by radial drift, at different wavelengths we would observe truncation at different radii since radial drift affects different grain sizes with different efficiency. This is not what was measured by \cite{Tazzari21}, suggesting that the outer edge of the continuum emission is caused by a drop in the dust density, and not by an opacity feature.

While \cite{Tazzari21} conclude that observations of millimeter continuum emission are not enough to distinguish which scenario is most likely in IM~Lup, \cite{Avenhaus18} are more specific and identify a outer edge at 340~au in the dust scattered light emission. This structure could trap grains at its location, and the grains could follow a different distribution inside and outside the ring. In our model, we recover the small grain distribution from the large grain distribution using the power law distribution in Eq.\ref{eq:size_distribution}. However, if the grain distribution changes at the two sides of the ring, and there is no large grain emission in the outer disk, we cannot recover the small grain abundance using this approach. Therefore in our fitting routine we do not include the region outside the ring location, at 340~au.




\section{Inferred disk model parameters}
Starting from the geometrical structure of IM~Lup, inferred directly from the observations as discussed in Sec.\ref{sec:observations}, and previous modeling efforts of its gas structure \citep{Zhang21}, we can put quantitative constraints on the disk physical parameters using the Markov chain Monte Carlo sampler \texttt{emcee}\footnote{https://emcee.readthedocs.io/en/stable/}. The physical distributions explored by the fit are:

\begin{itemize}
    \item The grain size distribution (Eq.\ref{eq:size_distribution}):
          \begin{equation}
              \label{eq:na_fit}
              n(a) \propto a^{-q},
          \end{equation}
          with the exponent of the distribution $q$ as fitting parameter.\\

    \item The maximum grain size radial distribution (Eq.\ref{eq:a_max}):
          \begin{equation}
              \label{eq:amax_fit}
              a_{max} = a_0 \; \left( \frac{r}{r_\mathrm{c}} \right)^{-p}
          \end{equation}
          with $r_c = 300$~au and $a_0$, $p$ as our fitting parameters.\\

    \item The dust-to-gas ratio radial distribution (Eq.\ref{eq:d2g}):
          \begin{equation}
              \label{eq:d2g_fit}
              \varepsilon(r) = \varepsilon_0 \; {\left( \frac{r}{r_\mathrm{in}} \right)}^{-\gamma}
          \end{equation}
          with $r_\mathrm{in} = 160$~au and $\varepsilon_0$, and $\gamma$ as our fitting parameters. We assume the dust-to-gas ratio. and the maximum grain size distribution to have the same $r_{cutoff}$.
    \item The turbulent viscosity parameter $\alpha_{turb}$, according to the standard \cite{Shakura73} viscosity recipe. This parameter sets the gas viscosity $\nu$, as shown in Eq.\ref{eq:diffusivity}. The turbulent diffusivity is used to compute the settling velocity and the turbulent diffusion coefficient of the dust species. These are used in turn to compute their settling mixing equilibrium, as discussed in Sec.\ref{sec:dust}.
\end{itemize}


This results in a model with 6 independent parameters \footnote{The source code can be found at:\\ https://github.com/rfranceschi/IMLup\_SPHERE\_fit}. The fitting is based on the definition of a reduced $\chi^2$. For the ALMA 1.25~mm emission, we measure the reduced $\chi^2$ between the observed emission profile and the azimuthal average of the synthetic millimeter images. Concerning the scattered light, as described in Sec.\ref{sec:observations_sphere}, we measure the average brightness profiles within four segments $10^\circ$ wide in PA along the major and minor axes (as in Fig.\ref{fig:obs_sca}). To give the same weight to the ALMA and SPHERE profiles, we consider the average $\chi^2$ of the scattered light  brightness profiles:

\begin{equation}
    \label{eq:chi2}
    \chi^2_{tot} = \chi^2_{mm} + \frac{1}{4} \left( \sum_{i=1}^4 \chi_{sca}^{(i)} \right)
\end{equation}

where $\chi_{sca}^{(i)}$ are the average brightness profiles in the four directions of the SPHERE images.

\begin{table}[]
    \centering
    \begin{tabular}{c c c}
        \hline\hline
        parameter       & value                              \\
        \hline          &                                    \\[-0.25cm]
        $q$             & $3.33_{-0.05}^{+0.03}$             \\[0.15cm]
        $p$        & $9.9_{-0.9}^{+0.6}$                \\[0.15cm]
        $a_0$           & $1.12_{-0.07}^{0.07} \times 10^{-2}$ cm       \\[0.15cm]
        $\varepsilon_0$ & $5.8_{-0.2}^{+0.5} \times 10^{-3}$ \\[0.15cm]
        $\gamma$        & $0.84_{-0.05}^{+0.08}$             \\[0.15cm]
        $\alpha_{turb}$ & $2.9_{-0.8}^{+0.5} \times 10^{-3}$ \\[0.15cm]

        \hline
    \end{tabular}
    \caption{IM~Lup best fit parameters of Eq.\ref{eq:na_fit}-\ref{eq:d2g_fit}. The fit values and the uncertainties are given by the 16th, 50th, and 84th percentiles of the samples in the marginalized distributions.}
    \label{tab:best_fit}
\end{table}

\begin{figure*}
    \centering
    \includegraphics[width=16cm]{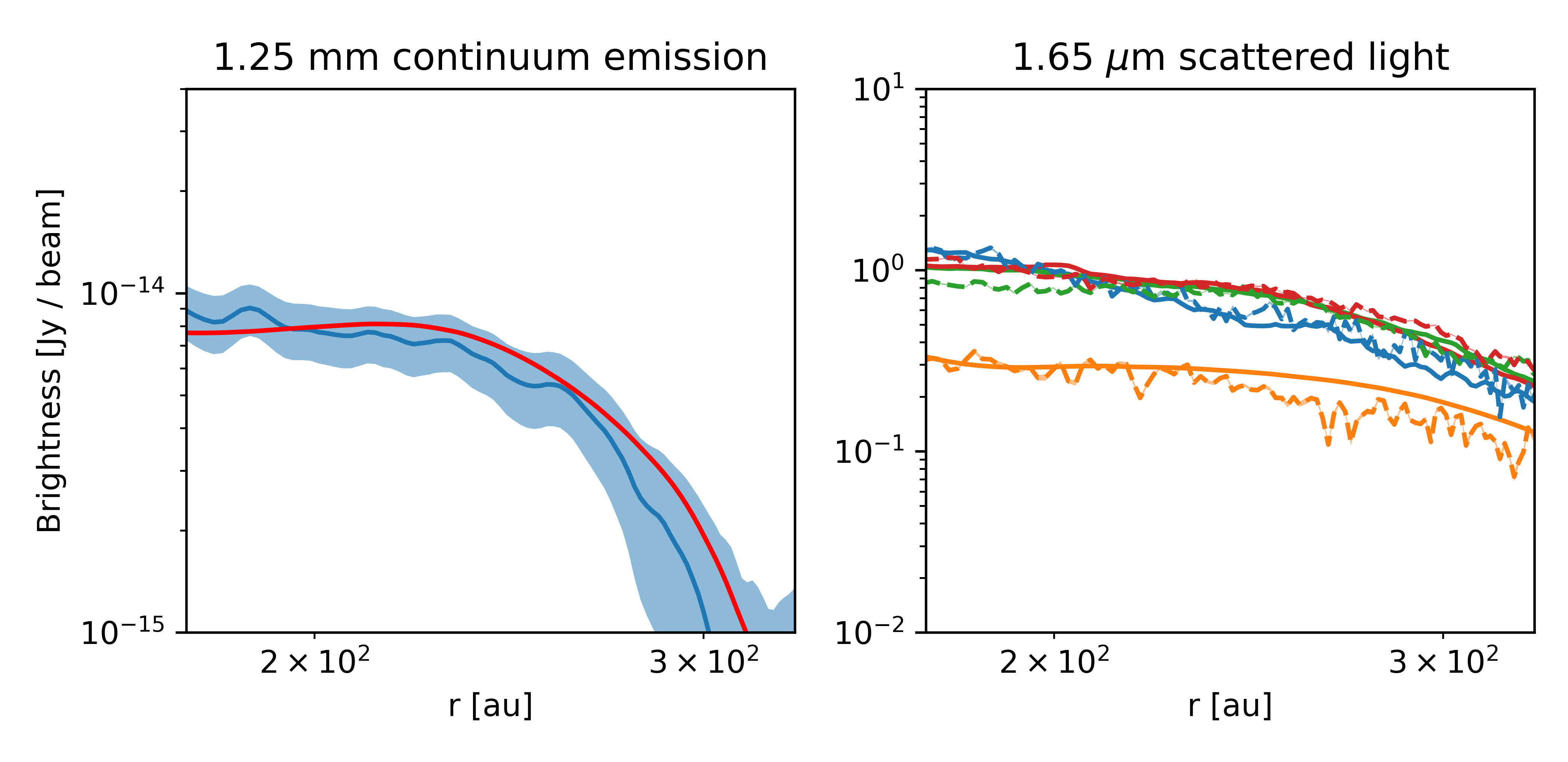}
    \caption{Comparison between the observation and best fit radial profiles for the millimeter continuum emission (left) and near-infrared scattered light emission (right). The shaded region shows the observation noise level. In the right plot, the four colors indicate the profiles in the four different directions, same as in Fig.\ref{fig:obs_sca}.}
    \label{fig:best_fit_profiles}
\end{figure*}

\begin{figure*}
    \centering
    \includegraphics[width=14cm]{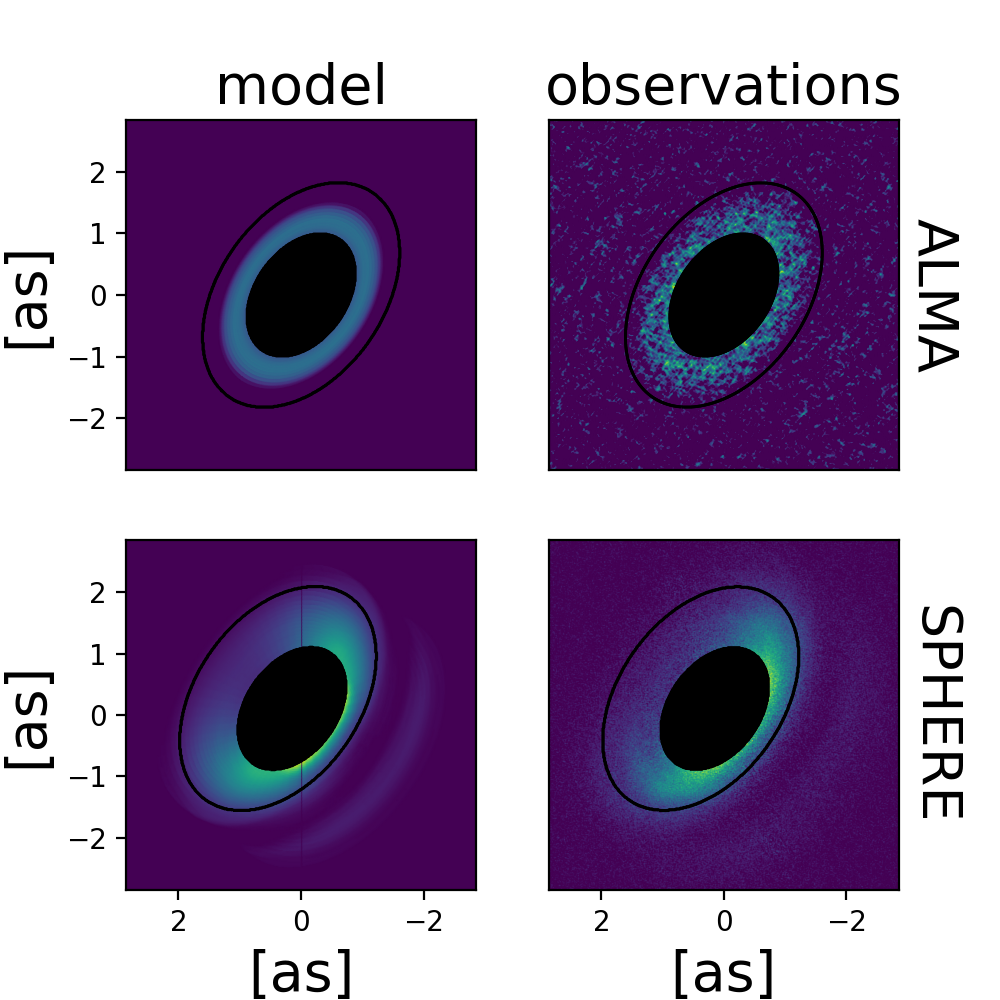}
    \caption{Comparison between the simulated 1.25~mm continuum and 1.65~$\mu$m polarized emission, and the ALMA and SPHERE data. The black masks show the region inside 1~arcsec radius, not included in our fitting, where radial structures are observed. The black circle is the 2.5~arcsec radius, the outer edge of the disk.}
    \label{fig:images comparison}
\end{figure*}

The best fit parameters are reported in Tab.\ref{tab:best_fit}, and the comparison between the synthetic and observed profiles is shown in Fig.\ref{fig:best_fit_profiles}. In Fig.\ref{fig:images comparison} we compare the simulated images to the observations. The best fit model well reproduces both the ALMA and SPHERE images. The simulated emission profiles at both wavelengths (and the four directions for the scattered light images) match the observations. Models and observations are less in accord in the inner disk ($< 1$~arcsec). As discussed in Sec.\ref{sec:observations}, the inner disk shows evidence of structures which we are not including in our models, and the coronograph used in SPHERE observations masks the emission in the very inner disk of the scattered light image. Our 1+1D disk model, as discussed in Sec.\ref{sec:model}, does not take into account radial transfer of energy, and the differences between model and the data in the inner disk structure is within the model uncertainty of using a 1+1D model rather than a full 2D model. Therefore, we do not include the profiles inside 1~arcsecond in our $\chi^2$ calculation, as they do not significantly affect our outer disk model.

\begin{figure*}
    \centering
    \includegraphics[width=14cm]{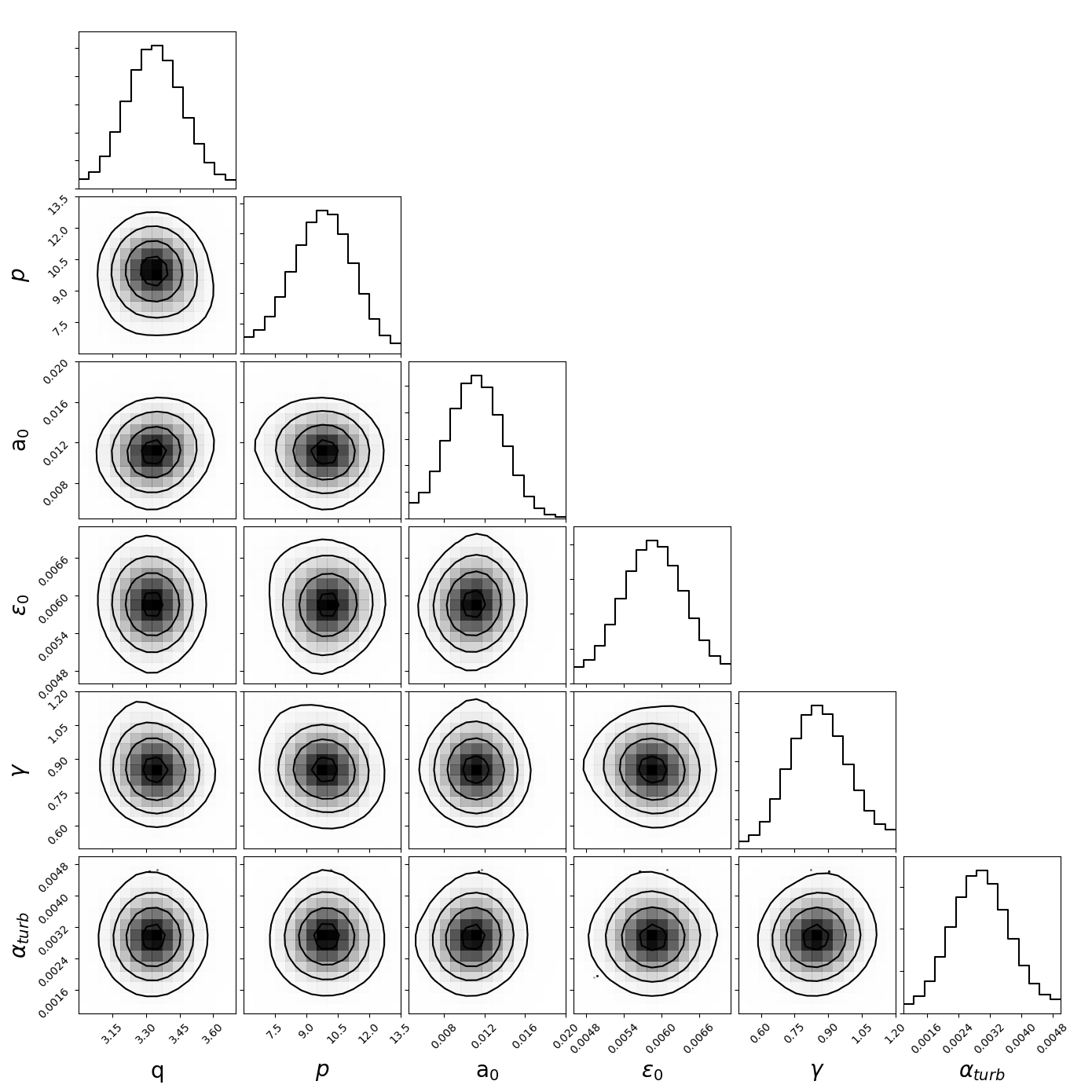}
    \caption{Posterior distribution of the fitting parameters described in Sec.\ref{sec:model}. The parameters follow a Gaussian distribution, showing that the parameters are well constrained and not correlated.}
    \label{fig:corner}
\end{figure*}


\subsection{Dust mass}
\label{sec:disk mass}
A key result of our modeling is the constraints on the dust mass of the disk. In the optically thin case, the millimeter flux is proportional to the disk dust mass times the dust opacity. The opacity at millimeter wavelengths is the highest for millimeter-sized grains, but it is only a factor of a few higher than the millimeter opacity of submillimeter-sized grains (e.g. \citealt{Woitke16, Birnstiel18}, also Fig.\ref{fig:optical_constants}). If millimeter grains are absent, we can still observe the millimeter emission from submillimeter-sized grains. However, dust mass estimates from the continuum millimeter emission are usually based on the assumption that the emission is coming from millimeter-sized grains. This assumption however cannot be justified by looking only at the millimeter emission. Our study takes into account both the emission from small and large grains, without assuming the size of the emitting grains. However, the reader should keep in mind that our grain distribution is constrained by the observations only in the region between 160~au and 340~au, as explained in Sec.\ref{sec:observations_sphere}-\ref{sec:millimeter observations}, and our dust mass estimate can be off because the dust distribution in the inner disk is extrapolated from the best-fit dust distribution in the 160-340~au region.

The gas mass is fixed at $\mathrm{0.2 \; M_\odot}$, derived from thermo-chemical modeling of ALMA high resolution CO emission maps \citep{Zhang21}. While it could be possible to find multiple grain size distributions reproducing the millimeter and near-infrared emission, the corner plot in Fig.\ref{fig:corner} shows us that this is not the case, as the posterior probability distribution converge to a unique solution. As a further proof that the presence of millimeter grains is necessary to simultaneously match the millimeter and near-infrared emission, we also run our fitting routine using a model where the grain size is not allowed to be larger than $10^{-3}$~cm. In this case, the disk model predicts a high abundance of small grains to match the millimeter emission, while still poorly reproducing the data. The large abundance of small grains causes the disk to be more flared than it appears the observations, suggesting that millimeter grains are required to reproduce the observed geometry.

As a further test, we check how well our model constrains the maximum grain size distribution by changing the $a_0$ parameter from Eq.\ref{eq:a_max} in our best-fit model. By halving $a_0$,  the millimeter image appears to be much less extended than the observations, since now millimeter-sized grains are less extended in the radial direction. The disk appears more flared in the scattered light, causing the disk surface in the forward direction to be more aligned to the observed line-of-sight. The scattered light polarization, in this direction, is at its minimum, and therefore the disk appears much dimmer in this region compared  to the observations. If we instead double $a_0$,  the millimeter emitting region is now more extended than in the observations, but also dimmer, since we lose some millimeter-sized grains to produce larger grain sizes. In the scattered light, the disk is now less flared than the best-fit model. This causes the light in the forward scattering direction to be more polarized, and here we observe a brightness excess compared to the observations. These effects of the maximum grain size distribution on both the millimeter and scattered-light profiles prove that this distribution is well constrained by our model, and that the presence of millimeter-sized grains is necessary to explain simultaneously both  the millimeter continuum and the near-infrared scattered light observations.

From the dust-to-gas ratio of our best fit model, we find a dust mass of $1.5 \times 10^{-4} \mathrm{M_\odot}$ in the fitted region. When we extrapolate our model results to the whole disk structure, we get a total dust mass of $\mathrm{4 \times 10^{-3} \; M_\odot}$. This dust mass estimate is about half the previous estimate of 0.01~M$_\odot$ by \cite{Pinte08}, based on 1.3~mm continuum emission obtained with the Submillimeter Array (SMA) \citep{Panic10}. This difference comes from the different assumptions made on the grain size distribution, as they assumed well-mixed dust populations with a fixed maximum grain size. They found that a maximum size of 3~mm and a disk dust mass of 0.01~M$_\odot$ reproduces the 1.3~mm emission, but not the silicate emission features from micron-sized grains observed in the mid-IR spectra. They suggest that to account for all observables a spatial dependence of the dust grain size distribution is necessary, with larger grains closer to the disk midplane and small grains on the disk surface. Indeed, our model features a vertical distribution of grain sizes, with a radial dependence of the maximum grain size. This results in a lower abundance of large grains in the outer disk. Since these grains carry most of the dust mass (see also the discussion in Sec.\ref{sec:distribution of the dust populations}), our model results in a lower dust mass than the one found in \cite{Pinte08}. Moreover, \cite{Pinte08} used an older estimate of the disk distance of 190~au, based on Hipparcos parallax measurement \citep{Wichmann98}, which is higher than the value used in this work, 158~au \citep{GAIA16}. In their model, the emission of large grains (which carry most of the dust mass) is then brighter by a factor of about 1.4, leading to an overestimation of the dust mass of the same factor. 

\begin{figure}
    \centering
    \includegraphics[width=9cm]{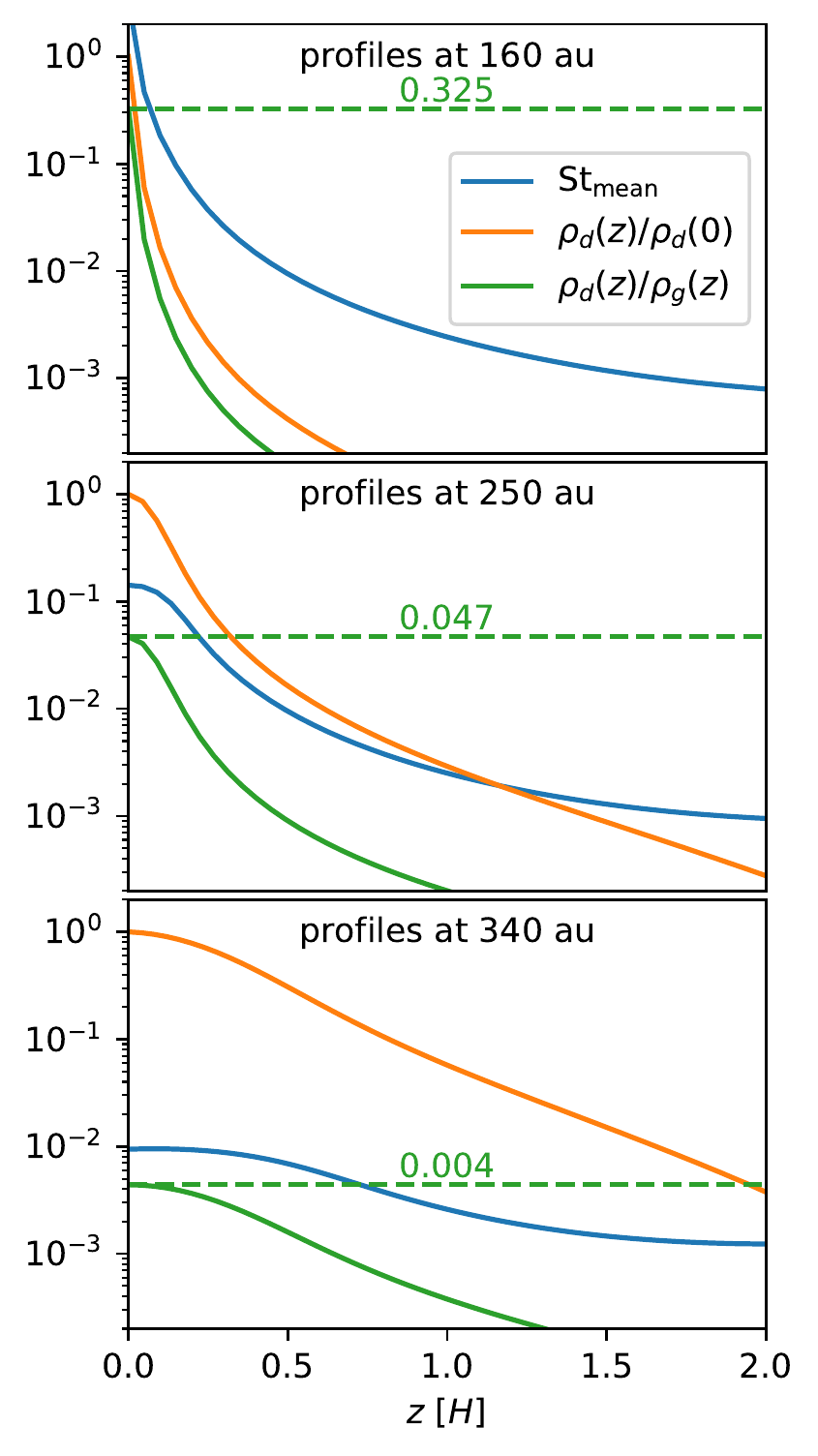}
    \caption{Stokes number, dust vertical profile, and dust-to-gas ratio at 160, 250 and 340~au radii (from top to bottom).}
    \label{fig:vertical profiles}
\end{figure}

To better understand the best-fit dust distribution we look at the vertical distribution of the dust properties, and how they change with the radial location. In Fig.\ref{fig:vertical profiles} we show the Stokes number, the dust vertical profile $\rho_d(z) / \rho_d(0)$ and the vertical dust-to-gas ratio $\rho_d(z) / \rho_g(z)$. The Stokes number is defined as the grain stopping timescale times the Keplerian frequency: $St = \tau_{stop} \; \Omega_K$, and it describes how fast the grains drift towards the inner disk. At smaller radii, we have a high Stokes number close to the midplane. This is due to large grains that are quickly drifting towards the inner disk, while at larger radii the midplane has already been depleted from the larger grains by drift. Above the midplane, we have a low Stokes number at all radii. This is caused to the vertical sedimentation of larger grains, suggesting that the vertical sedimentation happens on a much faster rate than the radial drift.

At different radii we find significant differences in the dust-to-gas ratio vertical profiles. While at larger radii the profiles get flatter, with a overall lower ratio, at smaller radii the dust-to-gas ratio is strongly peaked at the mid-plane, where it gets as high as 0.3. This is an important result, as at such high dust-to-gas ratios streaming instabilities can start to develop, possibly setting up the right conditions for planetesimal formation \citep{Youdin2005,Johansen2009,Bai2010}. Moreover, while not included in our disk model, observations show evidence of structures in the disk midplane inside the 160~au radius (as discussed in Sec.\ref{sec:millimeter observations}). If there are indeed active planet formation processes in the inner disk, the growing planets could then imprint non-axisymmetrical structures (e.g. \citealt{Lovelace99, Li00}). Our model, independently from this observational evidence, suggests that at this radius we find the right condition for the developing of streaming instabilities, possibly triggering planet-formation processes which in turn could be the origin of the structures observed in the inner disk. 


\subsection{Distribution of the dust populations}
\label{sec:distribution of the dust populations}
\begin{figure}
    \centering
    \includegraphics[width=9cm]{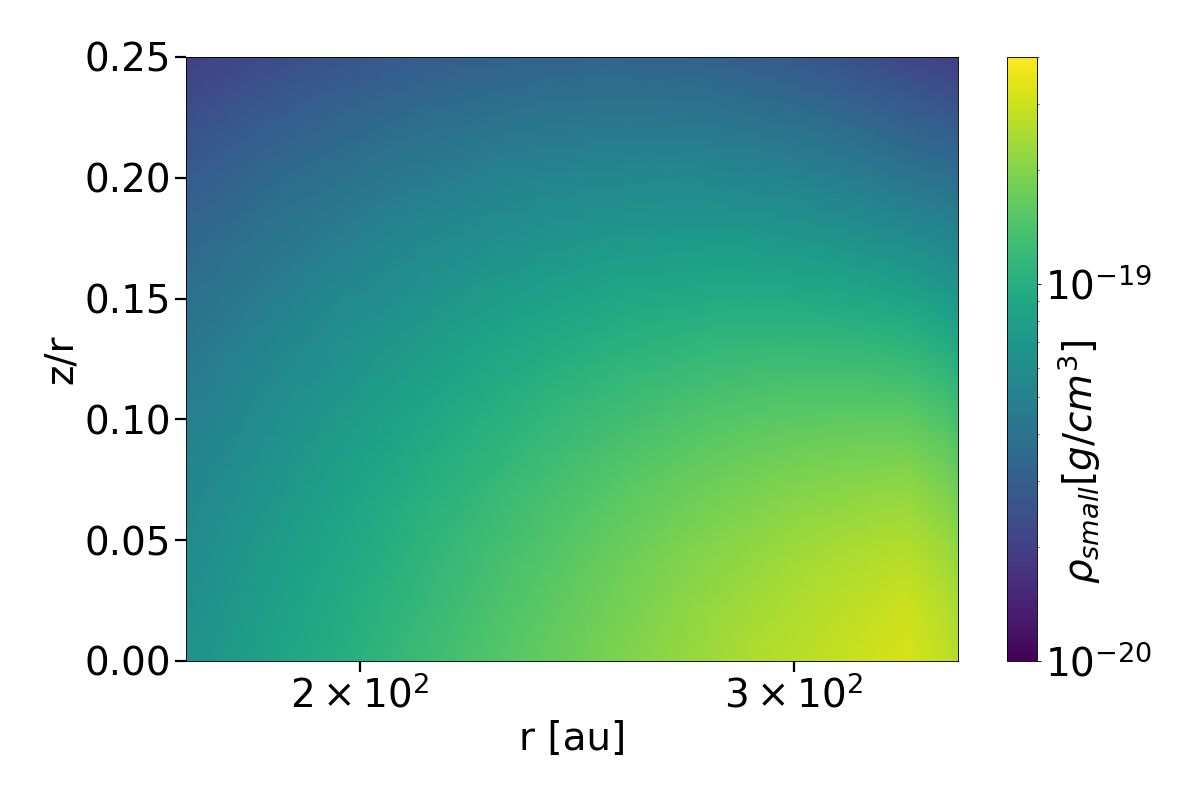}
    \caption{Density distribution of micrometric grains ($a_{grain} < 10^{-3}$~cm). These grains are more abundant in the regions where large grains are removed by radial drift (outer disk) and vertical settling (disk surface), and depleted in the midplane at smaller radii by dust growth processes.}
    \label{fig:small grain distribution}
\end{figure}

The dust mass distribution of different grain populations is an important topic for the understanding of dust evolution and planet formation. Our best fit model is in agreement with a vertical and radial stratification of dust grains. We find micron-sized grains on the disk surface ($z/r \approx 0.2$) and in the outer disk. They are however missing in the midplane at the smaller radii, where they can efficiently grow to form larger grains, as shown in Fig.\ref{fig:small grain distribution}.

\begin{figure}
    \centering
    \includegraphics[width=9cm]{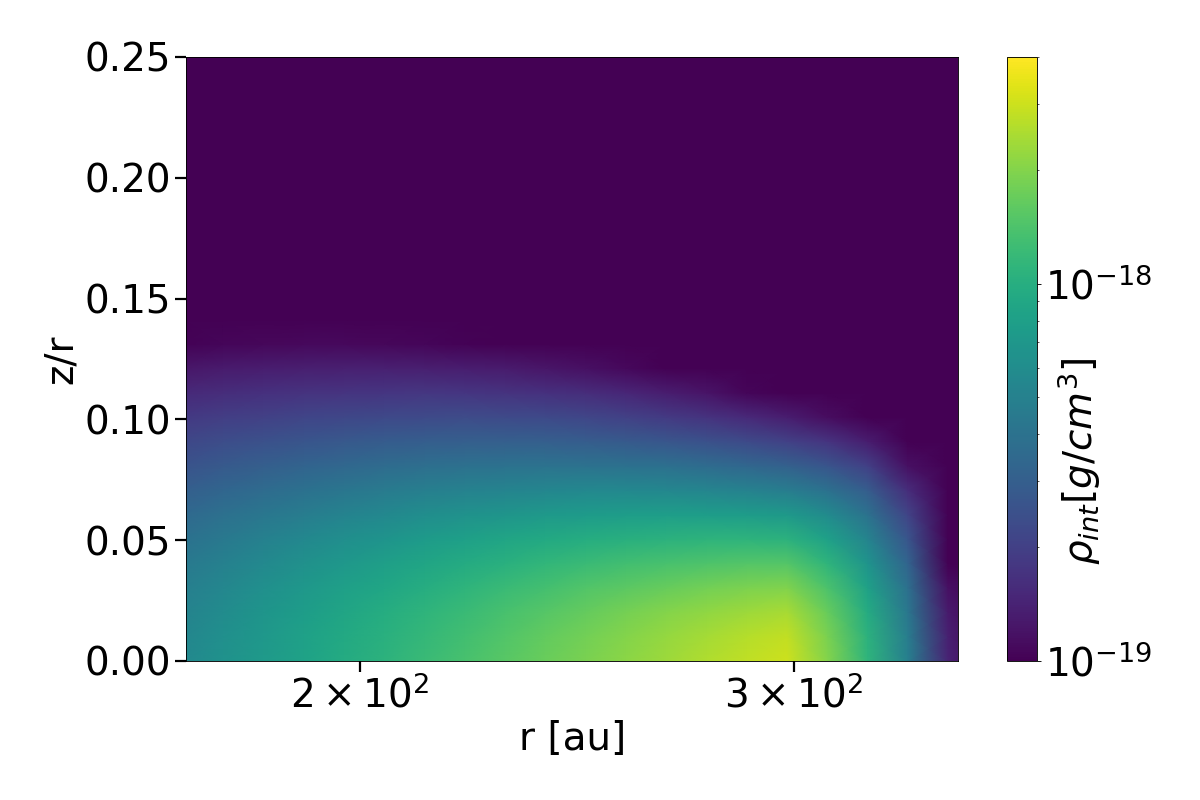}
    \caption{Density distribution of submillimeter grains ($10^{-3}$~cm~$ < a_{grain} < 10^{-1}$~cm). These grains have a moderate vertical extension, and are depleted at both small and large radii, where they are removed by dust evolution processes.}
    \label{fig:intermediate grain distribution}
\end{figure}

Submillimeter grains (Fig.\ref{fig:intermediate grain distribution}) are more abundant than micrometric grains, and are likewise depleted at smaller radii by the grain growth process. They also extend vertically in the disk, though they do not reach the disk surface ($z/r \approx 0.1$). As these grains are a product of dust evolution, they are not found in the outer disk, where they are removed by radial drift.

\begin{figure}
    \centering
    \includegraphics[width=9cm]{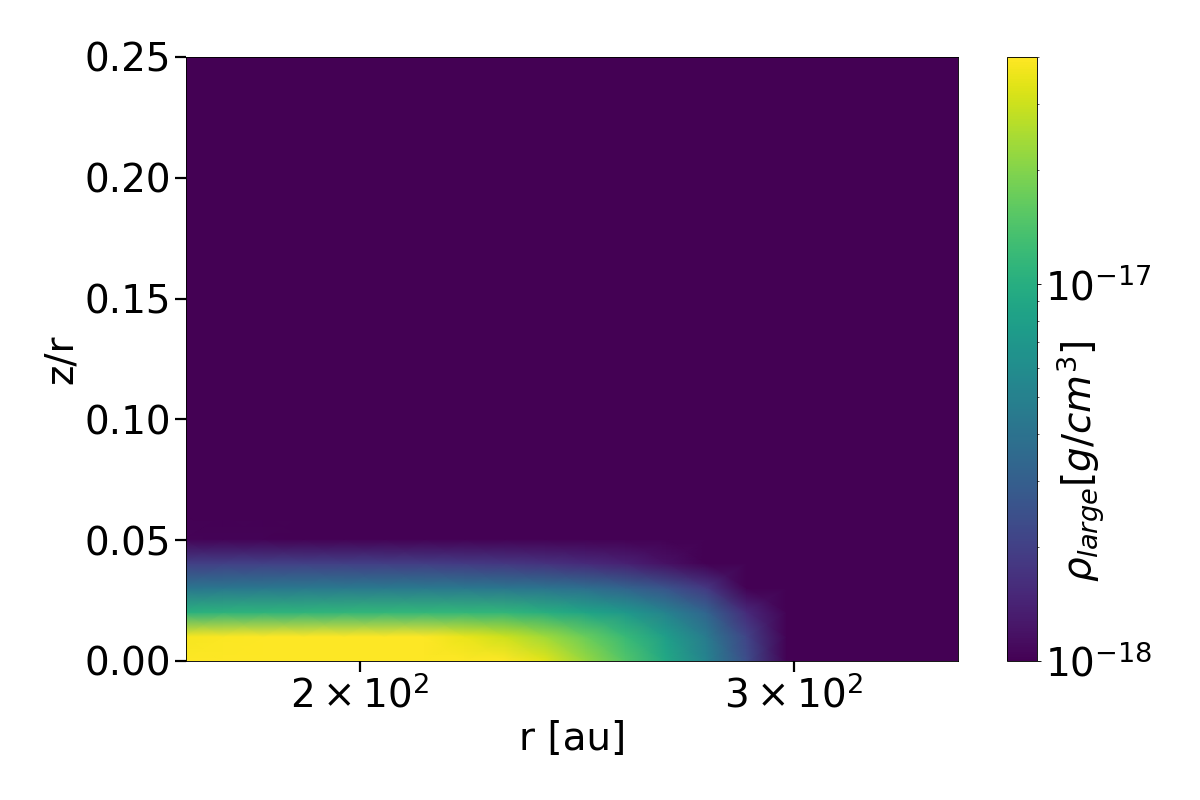}
    \caption{Density distribution of millimeter grains ($a_{grain} > 10^{-1}$~cm). These grains have a moderate vertical extension, and are depleted at both small and large radii, where they are removed by dust evolution processes.}
    \label{fig:large grain distribution}
\end{figure}

Millimeter-sized grains (Fig.\ref{fig:large grain distribution}), instead, are found exclusively in the midplane. In contrast to smaller grains, their density increases at smaller radii, as they are a  product of dust evolution which here happens on shorter timescales. They are more affected by radial drift, therefore they are less extended in radius than smaller grains.

\begin{figure}
    \centering
    \includegraphics[width=9cm]{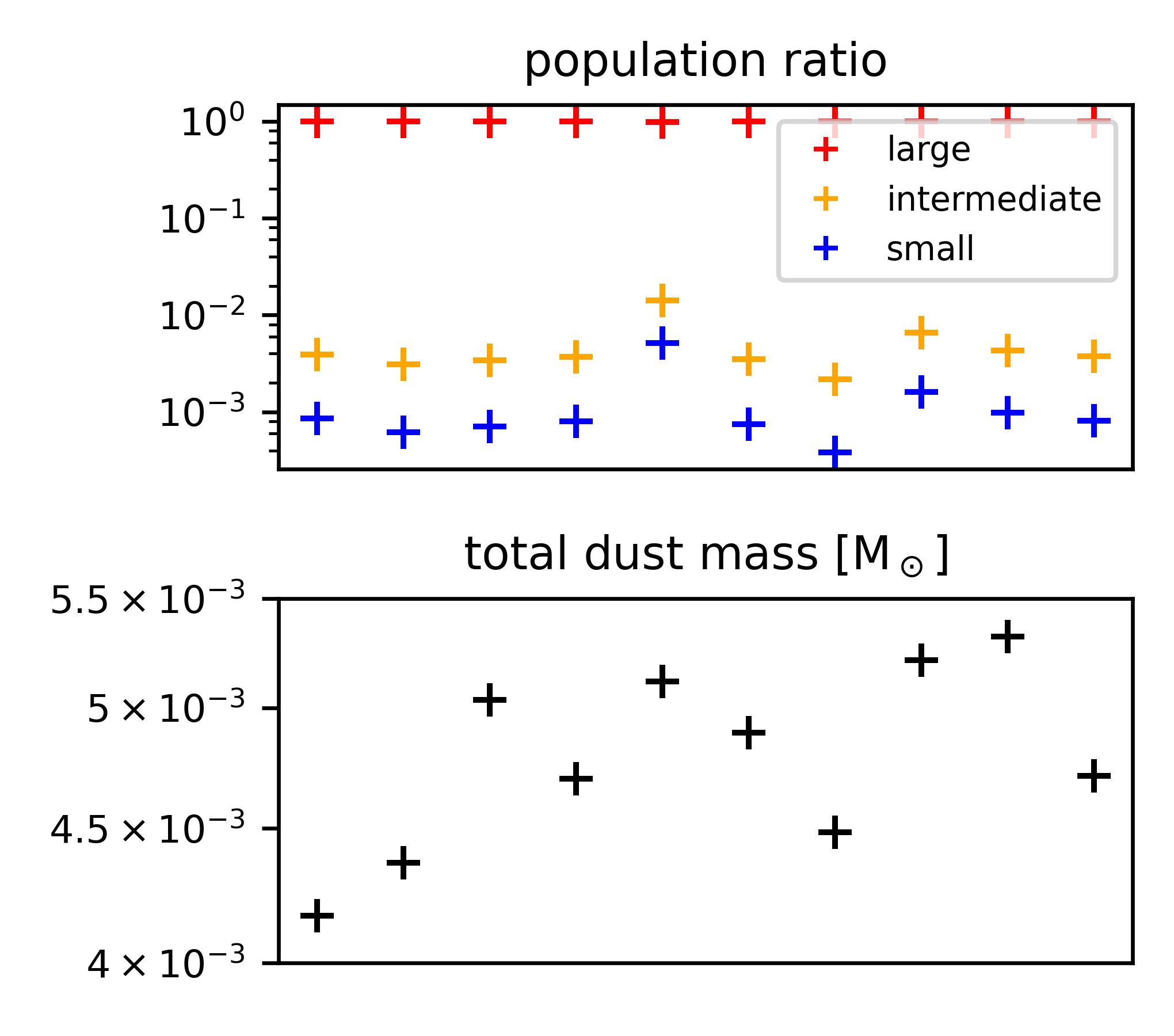}
    \caption{Upper panel: the mass ratios of large ($a_{grain} > 10^{-2}$~cm), intermediate ($10^{-3} < a_{grain} < 10^{-2}$~cm), and small grains ($a_{grain} < 10^{-3}$~cm) of the 10 best fit models. }
    \label{fig:mass ratios}
\end{figure}

The total dust mass, and the mass fraction in small, intermediate, and large grains, is tightly constrained by our fit. In Fig.\ref{fig:mass ratios} we show the mass ratio of each dust population to the total dust mass, and the total dust mass of the 10 highest likelihood models. The ratios and the total dust mass are in accord within all the models, proving that they are well constrained by observing both large and small grain emission.

In Sec.\ref{sec:disk mass} we discussed how the millimeter emission does not necessary trace the emission of millimeter-sized grains, as it is usually assumed. It is then worthwhile to check if, according to our model, this is a good assumption for the IM~Lup disk. The best fit grain size distribution, in Fig.\ref{fig:grain_distribution}, shows that, according to our model, the millimeter-sized distribution and the observed millimeter continuum emission overlap, and therefore the millimeter grains dominate the emission.

\subsection{Grain evolution timescales}
\begin{figure}
    \centering
    \includegraphics[width=9cm]{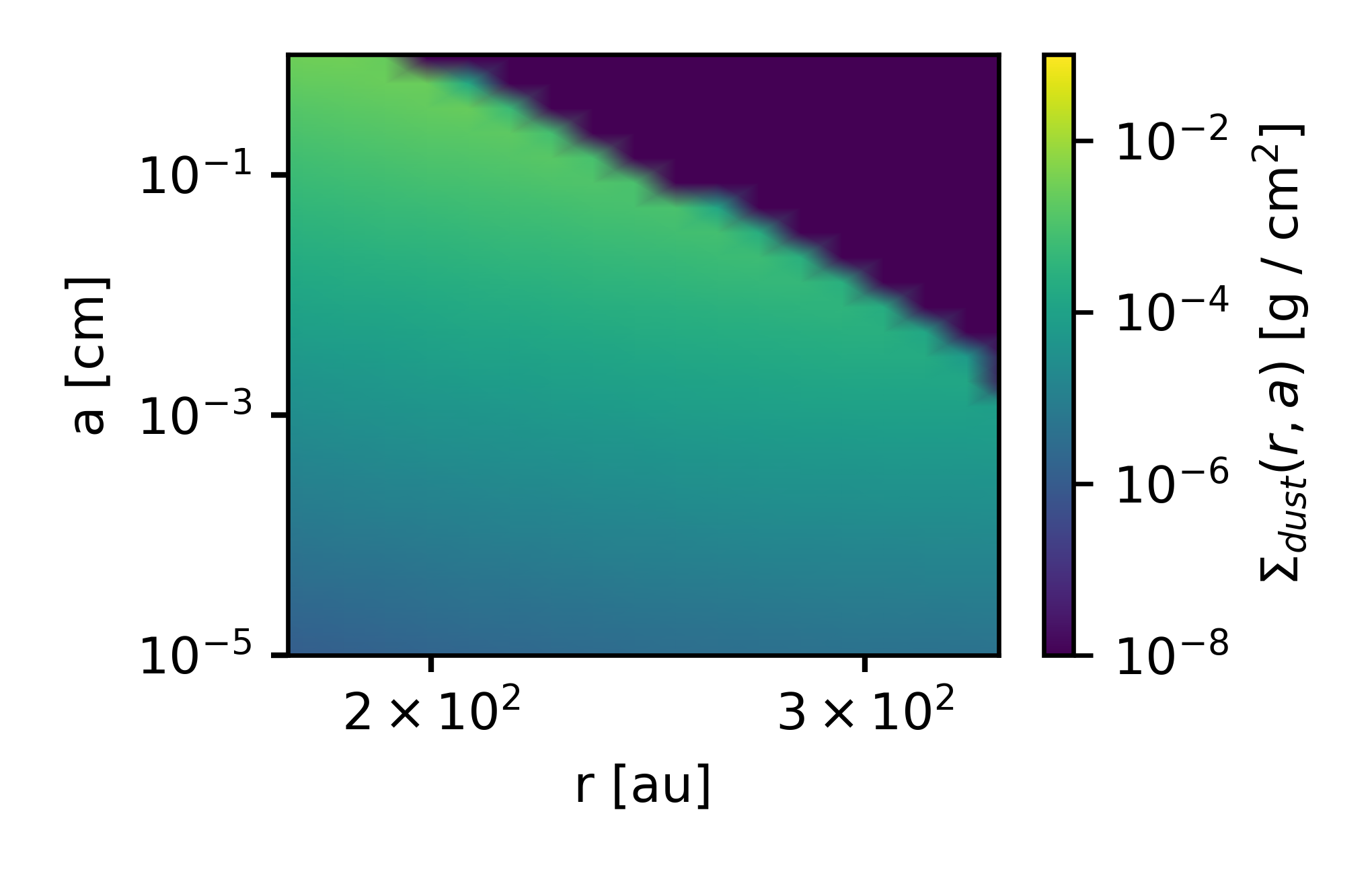}
    \caption{Dust surface density of the best fit model, as a function of grain size and radial position.}
    \label{fig:grain_distribution}
\end{figure}

Since our dust structure is the outcome of a parametric model, and not of a dust evolution simulation, the grain distribution shown in Fig.\ref{fig:grain_distribution} may not be physically consistent with the disk gas structure.  Moreover, our model suggests a steep maximum grain size distribution, as shown both in Fig.\ref{fig:corner} and Fig.\ref{fig:grain_distribution}. The parametrization assumed for the maximum grain size distribution, Eq.\ref{eq:a_max}, does not hold for all grain sizes. Once grains reach centimeter sizes, they break each other apart by colliding, limiting the maximum grain size to 1~mm - 1~cm, depending on the turbulence strength and gas density (e.g. \citealt{Birnstiel11, Zsom11}). This prevents grains to grow to unphysically large sizes in the region inside 1~arcsecond, which we do not include in our fit due to the presence of radial structures, as discussed in \ref{sec:millimeter observations}. However, it is worthwhile to investigate if the resulting grain distribution is physically consistent.

\begin{figure}
    \centering
    \includegraphics[width=9cm]{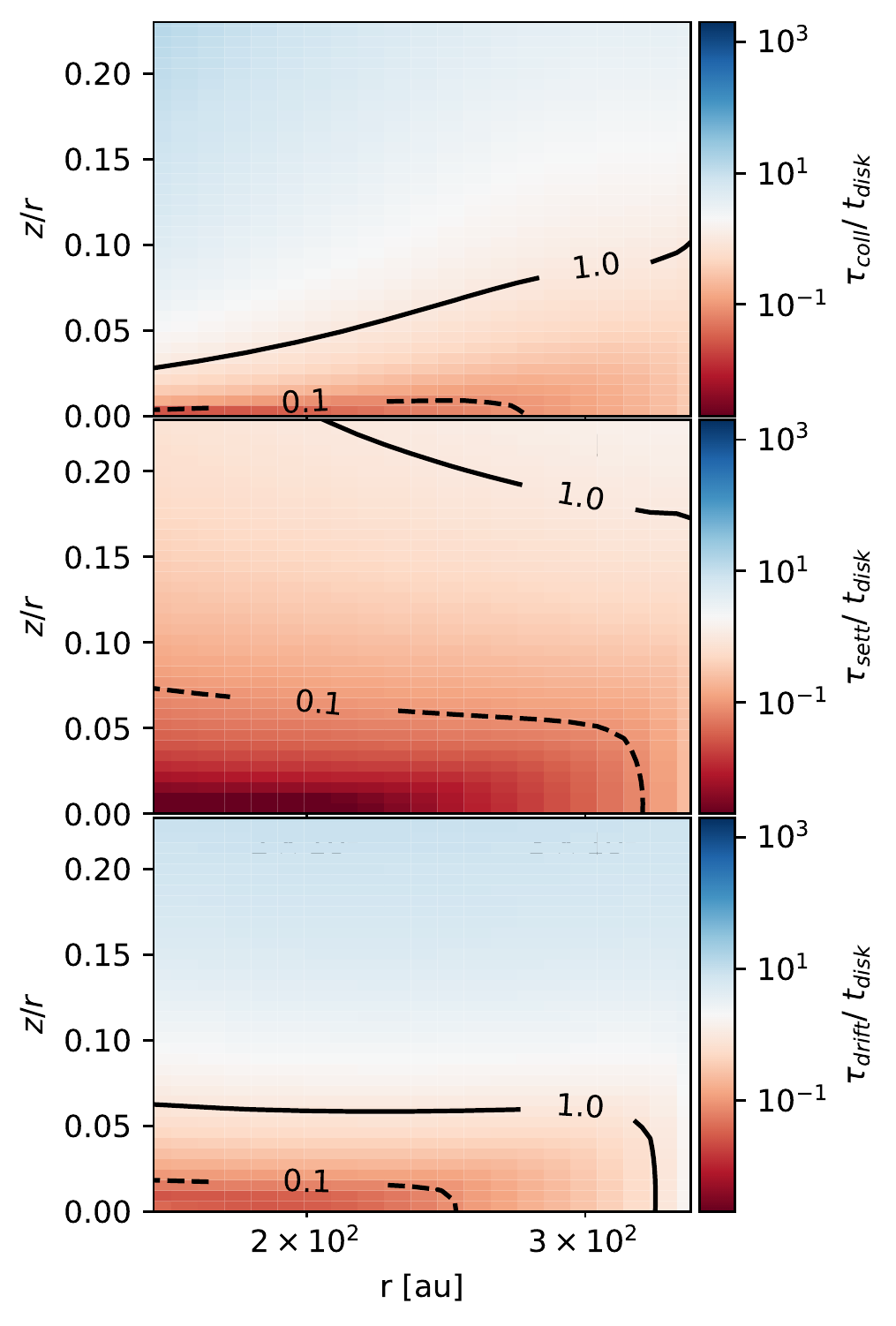}
    \caption{Comparison of the collisional, settling, and drift timescale to the disk age, from top to bottom.}
    \label{fig:timescales}
\end{figure}

In Fig.\ref{fig:timescales} we compare the collisional, settling, and drift timescales to the disk age, averaged over the grain size distribution. With the exception of the disk midplane, where grains can efficiently grow, the collisional timescale is much longer than the disk age. Here grains do not have enough time to grow through collision processes, and they cannot produce grains larger than micron sizes. However, our model predicts the presence of submillimeter grains even outside of the midplane, which then must be produced by other processes. A possible explanation is coagulation driven by sedimentation \citep{Zsom11}. To understand this process we must first discuss the grain settling timescales. Fig.\ref{fig:timescales} shows that small grains have a settling timescale of about the disk age on the disk surface, and did not have enough time to start settling to the midplane. Deeper in the disk the settling timescale gets shorter and grains start to settle towards the midplane. The difference in the settling velocity of grains drives their coagulation, as explored by several authors (e.g., \citealt{Dullemond04, Dullemond05, Schrapler04}). \cite{Zsom11} show that coagulation driven by sedimentation can be a very efficient process for grain growth, and can produce millimeter-sized grains as early as after $10^4$~orbital timescales ($10^4$ years at 1~au). Since the age estimates of IM~Lup range from 0.5~Myr \citep{Andrews18} to 1~Myr \citep{Andrews18}, this provides enough time for the formation of the large grains predicted by our model. The distribution of submillimeter grains shown in Fig.\ref{fig:intermediate grain distribution} is right inside the region where the settling timescale is shorter than the age of the disk, while the millimeter grain distribution corresponds to the region where the settling timescale is 10 times shorter than the age of the disk. This anti-correlation between the grain size and the low level of turbulence suggests that in the disk grain growth may be driven by sedimentation instead of turbulent collisions. For a more thorough justification of our model grain distribution, however, we would need to use 2D dust evolution models, which we leave to a future work.

Finally, the drift happens on a much longer timescale than the disk age everywhere but in the midplane, while only being efficient at radii smaller than about 250~au. It is interesting to point out that these regions match the ones where submillimeter and millimeter grains are found (see Fig.\ref{fig:intermediate grain distribution}-\ref{fig:large grain distribution}). This is also in accord with the discussion in Sec.\ref{sec:disk mass}, where we show how the Stokes number gets higher at smaller radii.

\subsection{Viscous turbulence parameter}
Previous studies have constrained the value of $\alpha_{turb}$ to be about $\alpha_{turb} = 10^{-3}$, based on the observation of turbulent motions in the disks \citep{Hughes11, Flaherty17}. However, a more precise value is more complex to obtain as the relation between the turbulent motions and $\alpha_{turb}$ depends on the nature of the turbulence \citep{Cuzzi01}. Our best fit model is in agreement with these findings, with $\alpha_{turb} = 3 \times 10^{-3}$. This is lower than the previous measurement by \cite{Pinte18} ($\alpha_{turb} = 9 \times 10^{-3}$). In this work they fit a parametrical disk model to ALMA band 6 dust continuum emission and CO, $^{13}$CO, and C$^{18}$O line emission. The $\alpha_{turb}$ is then estimated from the vertical dust settling in their best-fit model.

\begin{figure}
    \centering
    \includegraphics[width=9cm]{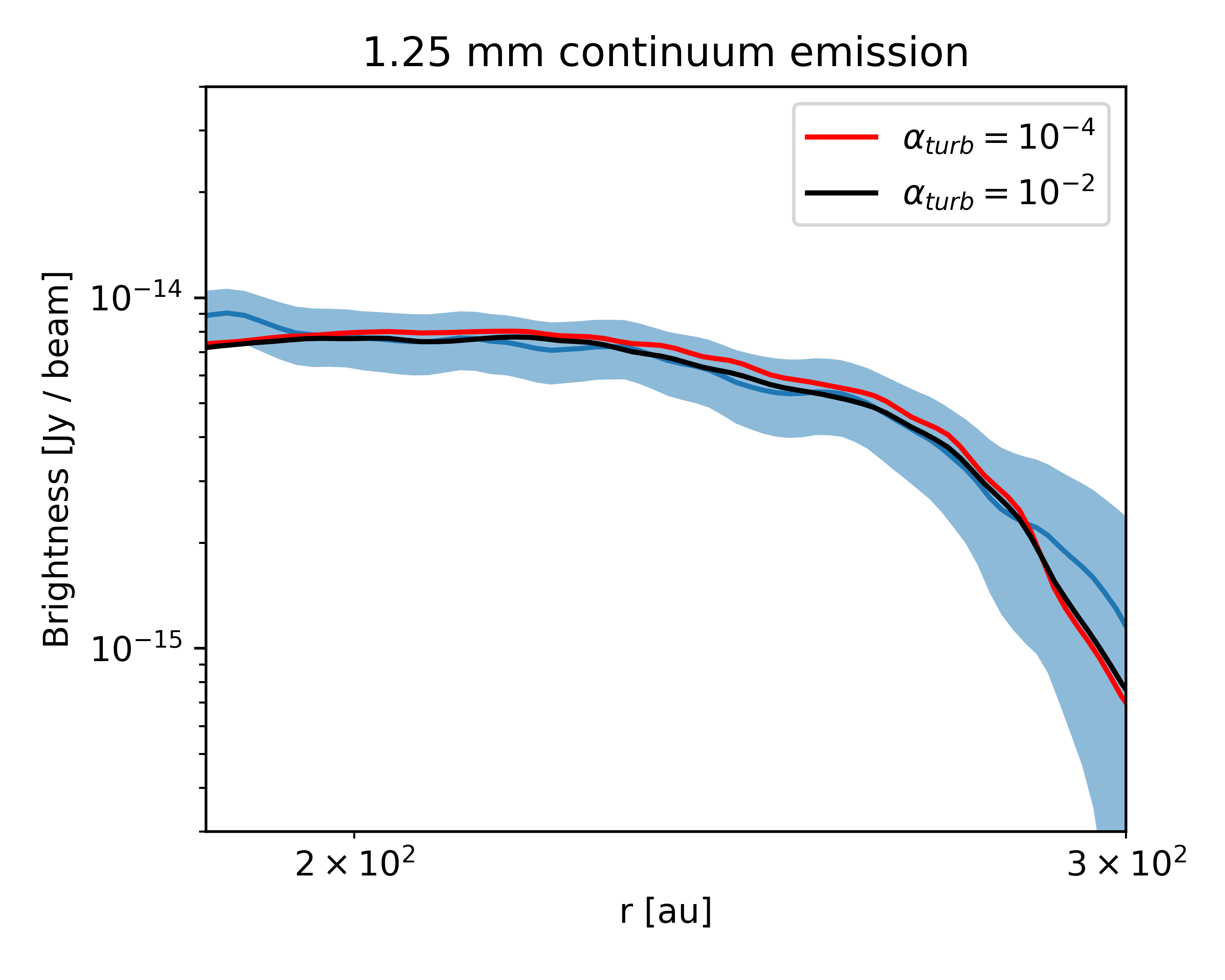}
    \caption{Comparison of the best-fit model millimeter emission profile but with low (red) and high turbulence (black) to the observed profile (blue).}
    \label{fig:alpha test}
\end{figure}

To check how the turbulence parameter affects the grain distribution, we run two models with the same parameters as our best-fit model, but with a low ($\alpha_{turb} = 10^{-4}$) and high ($\alpha_{turb} = 10^{-2}$) level of turbulence. In Fig.\ref{fig:alpha test} we show how the millimeter emission profiles of the new models compare to the profile of ALMA data. The millimeter profiles are unaffected by the change in $\alpha_{turb}$, and this is a further indication that the millimeter emission is coming from the large grains in the midplane, not impacted by turbulence.

The near-infrared emission, on the other hand, changes significantly with the turbulence strength. A high turbulent viscosity results in more grains in the disk atmosphere, as it prevents the dust from settling into the midplane. When $\alpha_{turb} = 10^{-2}$, the disk appears much more flared in the near-infrared, while when $\alpha_{turb} = 10^{-4}$ the near-infrared emission originates very close to the disk midplane, as the turbulence is not strong enough to sustain micron-sized grains.

This behaviour is also descrived by \cite{Rich21}. While they do not measure the value of $\alpha_{turb}$ (they assume a nominal value of $\alpha_{turb} = 10^{-3}$), they used an analytical model to manually explore the effect of $\alpha_{turb}$ on the $^{12}\mathrm{CO}$ emission height and small dust grain scattering surface. By decreasing the value of $\alpha_{turb}$, they found a lower scattering surface for small grains, in accord to our model prediction.




\section{Conclusions}
We present in this paper a model for the dust structure in IM~Lup reproducing both ALMA midplane millimeter continuum emission and SPHERE near-infrared scattered light emission from the disk surface. While these two wavelength are particularly sensitive to large (ALMA) and small (SPHERE) particles, we use these data to build a model for the full grain size distribution. The geometrical disk parameters are taken from the SPHERE DARTTS-S survey \citep{Avenhaus18}, and the gas mass from modeling CO isotopologues emission from ALMA data \citep{Zhang21}. The dust radial and vertical grain size distribution, and the dust-to-gas ratio, successfully reproduce ALMA millimeter continuum emission from the disk midplane \citep{Andrews18}, and SPHERE near-infrared scattered light emission from the disk surface \citep{Avenhaus18}.

Our main results can be summarized as follows:
\begin{enumerate}

    \item The posterior probability distribution of the model parameters shows a single solution for the dust distribution, which can reproduce the observations. We find that $\sim 99\%$ of the dust mass is carried by millimeter or larger grains, and a $10^{-4}$ mass fraction of micron-sized grains is sufficient to reproduce the scattered light emission surface observed by SPHERE. Assuming a gas mass of 0.2~M$_\odot$ inferred from CO emission \citep{Zhang21}, we estimate a dust mass of $4 \times 10^{-3} \; \mathrm{M_\odot}$.

    \item The turbulent parameter $\alpha_{turb}$ determines which particle sizes are sedimented and which are well mixed to the gas. We constrain the turbulence parameter to $3 \times 10^{-3}$, slightly higher than the usual assumed value $10^{-3}$. The turbulence strength does not affect the model millimeter emission, dominated by large grains unaffected by turbulence. On the other hand, the disk flaring observed in the near-infrared is strongly affected by the turbulence. At low values of $\alpha_{turb}$, the disk can appear flat also at small wavelengths, since small grains are free to settle to the midplane.

    \item Large millimeter-sized grains in the midplane are necessary to reproduce the near-infrared observations. It requires a vertical stratification of the dust populations, rather than being well mixed to the gas, with millimeter-sized grains in the disk midplane and micron-sized grains on the disk surface, in accord with dust coagulation models which include fragmentation (e.g. \citealt{Dullemond05, Birnstiel10}).

    \item The dust distribution shows a strong radial gradient in the particle size, a feature often associated with a pressure trap. It causes the disk to have a steep maximum grain size distribution, as the structures slow down the radial drift of the larger grains, which gather together on the inner side of the ring (e.g. \citealt{Pinilla15, Pinilla21}).

    \item We find a vertical  dust distribution with most of the mass carried by large grains in the disk midplane. The existence of large particles and the low turbulence in such a young disk point towards sedimentation driven coagulation.

    \item As a consequence of the strong radial gradient of the particle sizes distribution, we find millimeter and larger grains and a high dust-to-gas ratio in the disk midplane at smaller radii. It could be linked to the increase in brightness and the spiral structures observed in the inner disk, as in this environment dust drift is reduced, and streaming instability can start to develop, possibly triggering planetesimal formation.

\end{enumerate}


\begin{appendix}
    \section{Fitting the outer disk}
    In Sec.\ref{sec:millimeter observations} we discuss how the observations justify our assumption that the dust distribution changes the 340~au annulus, where a outer edge in the dust emission was observed \citep{Avenhaus18}. In our model, the abundance of small grain follows the one of the larger grains. However, if the dust distributions changes at the ring location, and there are no large grains in the outer disk, we cannot derive the abundance of small grains as the ring affects the large and the small grains differently. Another possible explanation for the sharp truncation in the millimeter profile could be an opacity feature. When the maximum grain size, decreasing with radius, drops below the one which emits most efficiently ($a = \lambda / 2 \pi$ according to the Mie scattering theory), we observe a sharp truncation of the emission profile. In our tests, we explored each of these two cases to check if they can reproduce the observations until the outer edge of the scattered light emission, at 400~au, instead of constraining our fit to the region inside the location of the ring, at 340~au.

\begin{figure}
        \includegraphics[width=9cm]{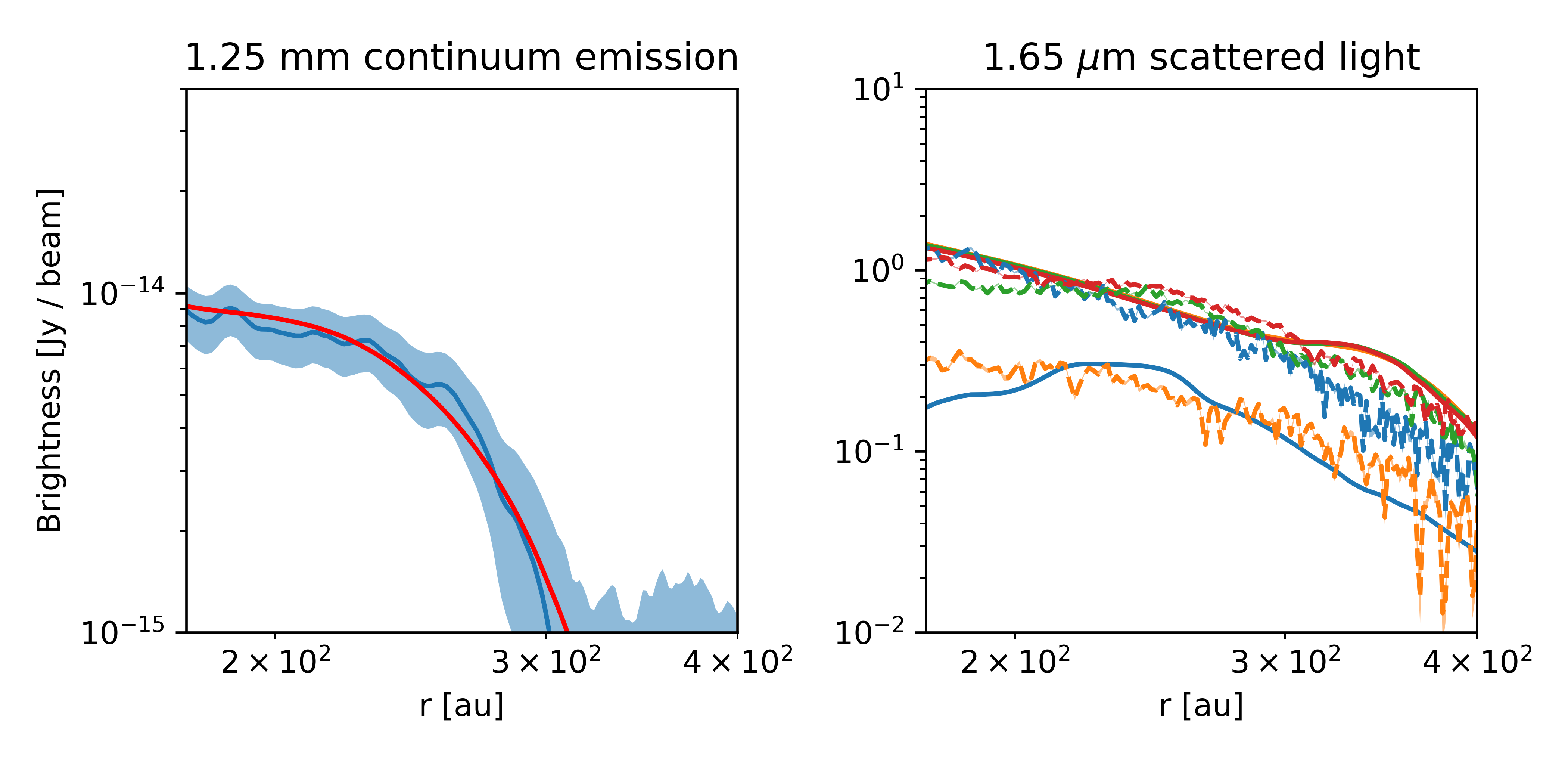}
        \caption{The model without a truncation in the dust distribution best reproducing the continuum data. This model does not include enough small grains to reproduce the scattered light observations.}
        \label{fig:model opac 1}
    \end{figure}

    To test if the observations can be explained by an opacity feature, we follow the same approach discussed in Sec.\ref{sec:model}, but we add to the fit the region outside the ring. This new model cannot reproduce both the continuum and the scattered light profiles. The model profiles which best fit the continuum emission are shown in Fig.\ref{fig:model opac 1}. The grain distribution reproducing both the brightness and the size of the continuum emitting region does not produce a small grain population which scatters light with enough efficiency to reproduce the scattered light emission.

\begin{figure}
        \includegraphics[width=9cm]{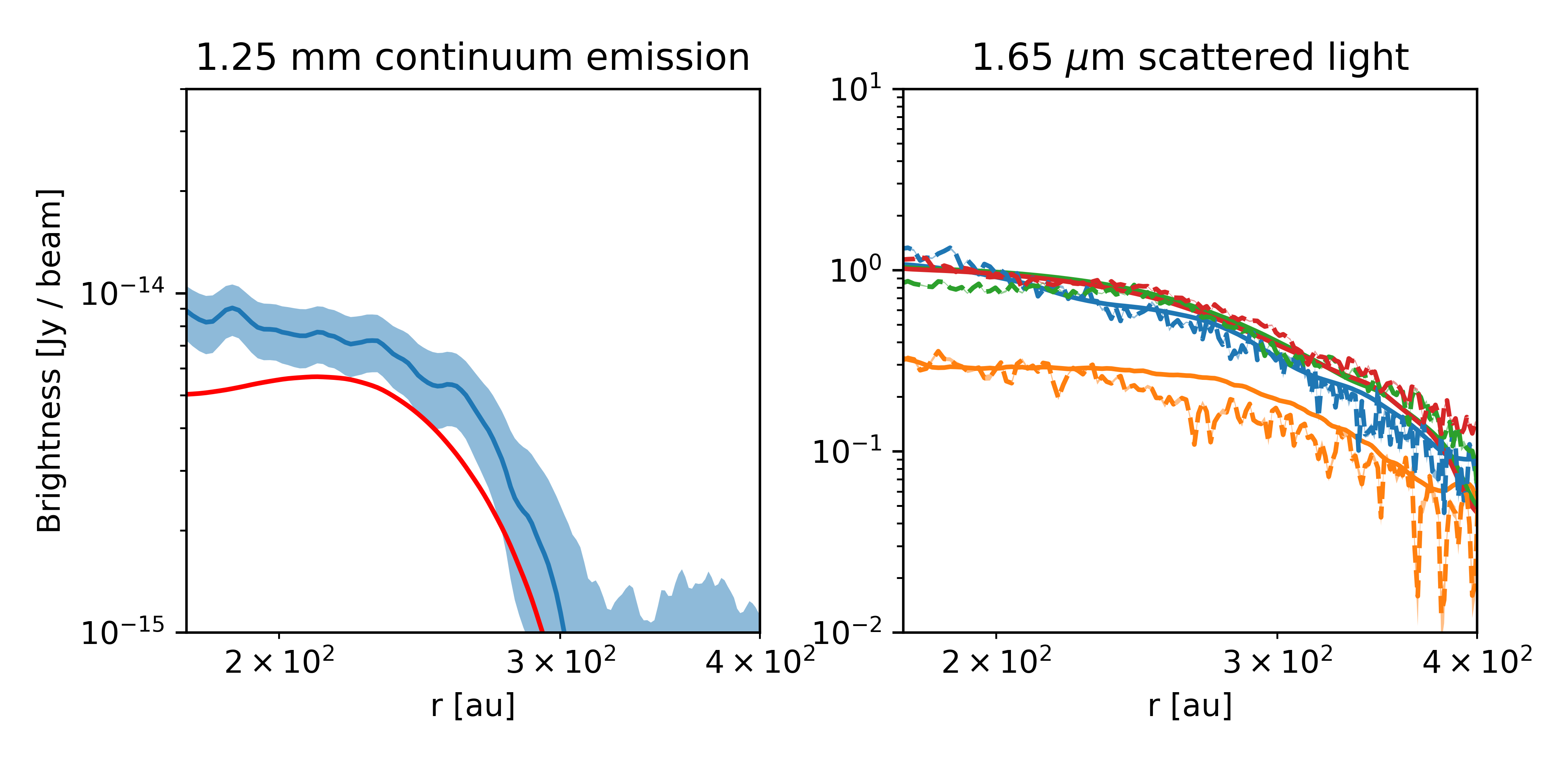}
        \caption{The model without a truncation in the dust distribution best reproducing the scattered light data. This model does not have enough millimeter grains to match the observed continuum emission.}
        \label{fig:model opac 2}
    \end{figure}

    The model best matching the scattered light data is shown in Fig.\ref{fig:model opac 2}.  This small grain distribution is coupled to an abundance of large grains which is too low to reproduce the continuum emission. Since this approach cannot reproduce at the same time the small and large grain emission, we conclude that the sharp drop in the continuum emission is not caused by an opacity effect, and a physical truncation of the dust distribution is necessary to explain the data.

\begin{figure}
        \includegraphics[width=9cm]{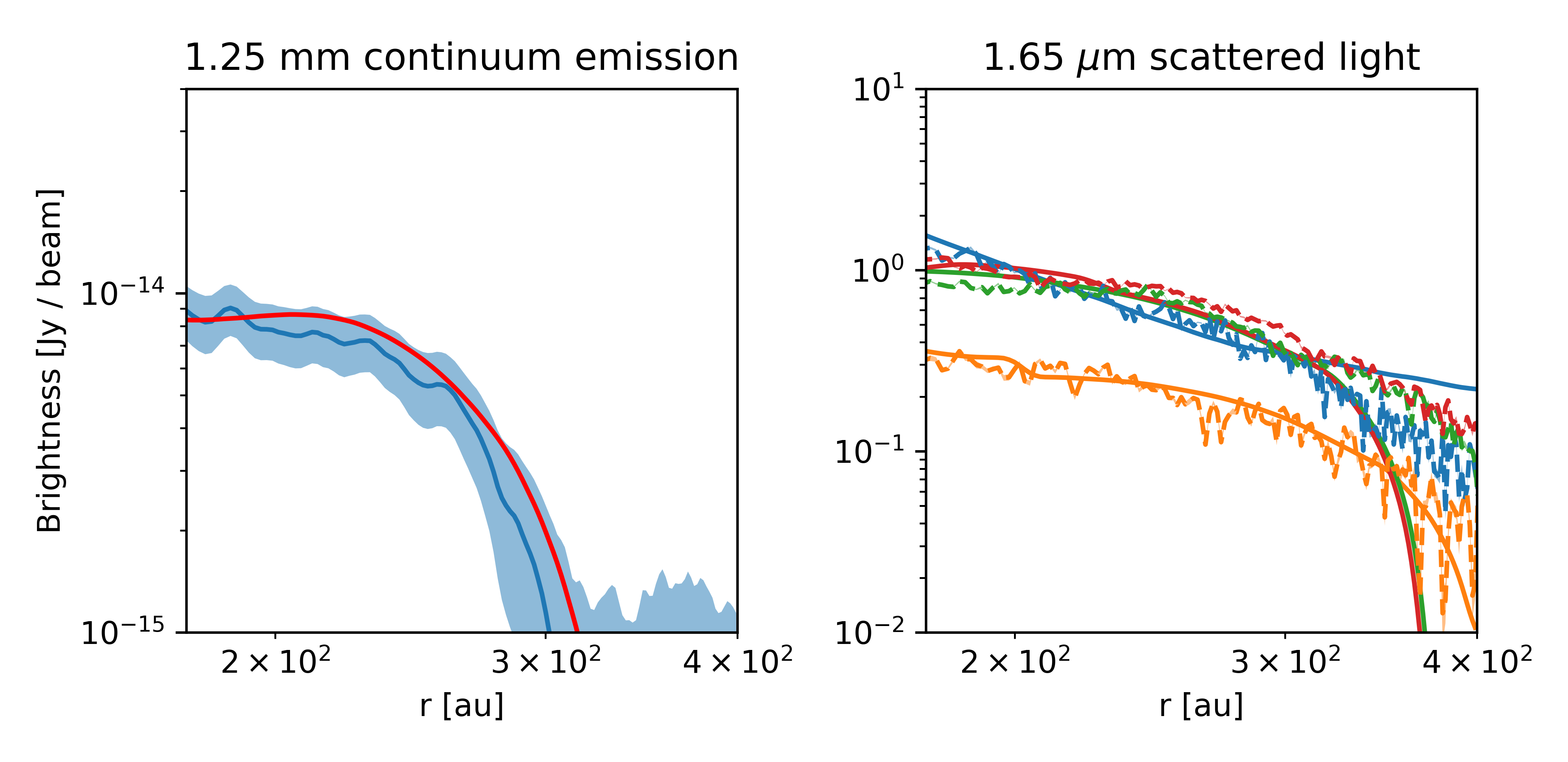}
        \caption{The best fit model with a truncation of the maximum grain size and dust-to-gas ratio distribution. This model well reproduces both the continuum and scattered light data in the region inside the truncation, but adds a sharp drop in the scattered light profiles not seen in the observations.}
        \label{fig:model cutoff}
    \end{figure}

    To test if a cutoff in the dust density can reproduce the observations, we modify Eq.\ref{eq:a_max} and Eq.\ref{eq:d2g} by adding an exponential cutoff:

    \begin{equation}
        a_{max} = a_0 \left( r / r_c \right)^{-\alpha} \; \exp \left( -r / r_{cutoff} \right)^{-p_1},
    \end{equation}

    \begin{equation}
        \varepsilon = \varepsilon_0 \left( r / r_{in} \right)^{-\alpha} \; \exp \left( -r / r_{cutoff} \right)^{-p_2},
    \end{equation}

    where $p_1$, $p_2$, and $r_{cutoff}$ are new fitting parameter, for a total of 8 parameters. The physical drop in the continuum emission at a specific wavelength can be explained by either a cutoff in the maximum grain size distribution, or in the dust-to-gas ratio. Since we do not have a theoretical argument to affirm in which distribution the cutoff is more physically justifiable, we allow for the possibility of a cutoff in both distributions, and let the fit to find out in which one fits the observed small and large grain emission. The profiles of best model produced by this approach is shown in Fig.\ref{fig:model cutoff}. This model places a cutoff of both the maximum grain size and  the dust to gas ratio at 300~au, the location of the outer edge of the continuum emission. This model well reproduces both the continuum and scattered light data, with the exception of the outer part of the disk. Here, the truncation removes the small grains emitting in the scattered light, and in the model we observe a sharp drop in the emission not present in the observations. While a truncation of the large grain distribution is necessary to reproduce the continuum observations, this would cause a similar truncation in the small grain distribution. This however is not what we observe in the data, because the ring in the dust structure affects the distribution of large grains, but not the one of small grains. Since there  are no large grains in the outer part of the disk, we cannot use our model to constrain the small grain distribution from the one of the large grains, and we constrain our fit to the region where the continuum emission indicates the presence of large grains.

\end{appendix}


\begin{acknowledgements}
    R.F. and T.H. acknowledge support from the European Research Council under the Horizon 2020 Framework Program via the ERC Advanced Grant Origins 83 24 28.\\
    We thank Myriam Benisty for the helpful discussions and for providing the SPHERE observational data.
\end{acknowledgements}

%
%

\bibliographystyle{aa} 
\bibliography{aanda.bib} 

\end{document}